C:/10comets/SLC L4+S1/V7/140601

# "The location of Oort Cloud Comets C/2011 L4 Panstarrs and C/2012 S1 ISON, on a Comets´ Evolutionary Diagram"

Ignacio Ferrín,
Institute of Physics,
Faculty of Exact and Natural Sciences,
University of Antioquia,
Medellín, Colombia, 05001000
ignacio.ferrin@udea.edu.co

Number of pages 56

Number of Figures 26

Number of Tables 4

**Abstract**

We plot the Secular Light Curves (SLCs) of comets C/2011 L4 Panstarrs and C/2012 S1 ISON. The brightness of C/2011 L4 Panstarrs and C/2012 S1 ISON increase steeply after discovery but both decrease in slope sharply at ~4-5 AU, a characteristic of Oort Cloud cometary light curves that we refer to as a Slope Discontinuity Event (SDE). After the SDE, the SLC of C/2011 L4 continues to increase and has a surge near perihelion. The SLC of comet C/2012 S1 ISON on the other hand, is very odd and exhibits a SDE plus a near-standstill after the event. We found five comets with similar behavior: C/1996 Q1 Tabur, C/1999 S4 LINEAR, C/2002 O4 Hönig, C/2010 X1 Elenin and C/2012 T5 Bressi, all of which disintegrated. Thus we predict that comet ISON will disintegrate too. We compiled published production rates of water, dust and CO and used them to calculate the Mass Loss Budget. We use this information for ISON to calculate the diameter and to plot 29 comets in an Evolutionary Diagram that separates comets by class. We find a diameter D = 1030±70 m in excellent agreement with the upper limit found by Delamere et al. (2013), D(mean) < 1126 m. It is evident that the Secular Light Curves, exhibit complexity beyond current scientific understanding. Note: The comet disintegrated as predicted (CBET 3731), while this paper was being refereed.

Key words:
comets: general, asteroids: general; Comet C/2012 S1 ISON

## 1. Introduction

The year 2013 offered the opportunity to observe two new comets coming from the Oort cloud, C/2011 L4 Panstarrs and C/2012 S1 ISON. Both have orbits with eccentricity e ~ 1.0. In this work we have reduced 16.673 photometric observations of eight comets, and arrive at significant scientific conclusions based on their Secular Light Curves (SLCs). The SLCs are new and have not been published previously. We also present plots of the temperature of the comets and their location on a Remaining Revolution vs Mass-Loss-age diagram, and a color-color diagram for comet ISON. Scientific data on these comets was obtained from the SAO/NASA Astrophysical Data System (ADS): http://adsabs.harvard.edu/abstract_service.html .

### 1.1 Comet C/2011 L4 Panstarrs

On June 6, 2011, astronomers at the Institute of Astronomy of the University of Hawaii, discovered a comet designed C/2011 L4 Panstarrs (Wainscoat et al. 2011).

Some results of interest for this investigation are due to Woodward et al. (2013) who measured an infrared temperature of 312 K at R= -0.84 AU, and Biver et al. (2012) and Opitom et al. (2013), who measured water production rates that will be used later on in Section 7.3 to calculate the water budget of this comet. Ivanova et al. (2014) conducted photometric observations from -4.4 to -4.2 AU, and Lovell and Howell (2013) made radio observations of OH.

### 1.2 Comet C/2012 S1 ISON

On September 21 of 2012, Nevsky and Novichonok (2012) discovered comet C/2012 S1 ISON at the notable distance of -6.3 AU. The object is coming from the Oort cloud with an original parameter 1/a = 0.000009, so the object is dynamically new. It had been increasing in brightness at a rate $R^{-5.02}$ and if it had continued at this rate, it would certainly have attained a magnitude much brighter than the full moon, which prompted many media reports to conclude that it was going to be “the comet of the century”, in spite of the fact that the century is just starting and we still have 87 years to go.

This work is based on measurement of the comet done by many authors. Schleicher (2013a, 2013b) measured the water production rate of the comet. Bodewits et al. (2013a) carried out photometric observations and were able to set an upper limit to the water production rate. Li et al. (2013) and Kelley et al. (2014) using the Hubble Space Telescope were able to set an upper limit to the diameter of the comet D < 4 km. Delamere et al. (2013) calculated 8 upper limits and find D(mean) < 1126 m. Lisse et al. (2013) made a series of measurements in different bands. Meech et al. (2013) made R-band measurements of the comet and compared them with a H2O, CO and CO2 production rate model. They conclude that the region around 5 AU (the SDE) was due to a slow CO outburst, a conclusion that is not supported by the current investigation. Bodewits et al. (2013 b) measured R magnitudes with a diaphragm of 10", with which they calculate dust production rates using the A'Hearn et al. (1984) formalism. Other authors include Sekanina (2013), Knight and Walsh (2013), and Hines et al. (2014). Sitko et al. (2013) measured the temperature of the comet at R= -0.69 AU, obtaining $T_{BB}$ = +335 $^{o}$K.

There has been a comet workshop on ISON with many results that have not yet reached the scientific literature. The workshop can be accessed at https://dnnpro.outer.jhuapl.edu/isonworkshop/Home.aspx .

The plan of this work is the following: In the Appendix we will list the data sets used in this investigation to plot the secular light curves (SLCs) described in Section 2. Section 3 is dedicated to comet C/2011 L4 Panstarrs and Section 4 to comet C/2012 S1 ISON. In Section 5 we study comets C/2002 O4 Hönig, C/1996 Q1 Tabur, C/1999, S4 LINEAR, C/2010 X1 Elenin, C/2012 T5 Bressi and C/1973 E1 Kohoutek and conclude that comet ISON will turn off or disintegrate and that comet Kohoutek did not really fizzle. Section 6 considers reasons to have a SDE. Section 7 defines de Mass Loss Budget and the Mass Loss Age. Section 8 introduces the *evolutionary diagram* Remaining Revolutions vs Mass-Loss Age, the final and most important result of this investigation. And Section 9 describes the Evolutionary Lines in that diagram.

In the following discussions, negative distances and negative log (distance) denote pre-perihelion measurements. Thus all distances and logs must be taken as positive.

## 2. Secular Light Curves (SLC)

### 2.1 Introduction

The SLCs give an unprecedented amount of information on the photometric behavior of comets, with more than 30 parameters measured from the plots (Ferrín 2005a=Paper I, 2005b, 2006, 2007, 2008, 2010a = The Atlas, 2010 b, 2012, 2013, 2014).

The protocol of the Secular Light Curves (SLCs) follows closely the procedures described in the *Atlas of Secular Light Curves of Comets* (Paper II). The preferred phase space to describe the luminous behavior is the $m_{COLOR}(1,R) = m_{COLOR}(\Delta,R) - 5 \log \Delta$ versus Log R plane, where $\Delta$ is the Comet-Earth distance and R the heliocentric distance of the comet. In the $m_{COLOR}(1,R)$ versus Log R diagram, powers of R, $R^{-n}$, plot as straight lines of slope 2.5 n. The $R^{-n}$ behavior is easy to recognize and measure. At the bottom of the plot, the nucleus magnitude appears as straight lines in the form of a pyramid, with a power law $R^{-2}$. This is the law that would exhibit an asteroid or an atmosphereless body.

To carry out this investigation we reduced 16.673 photometric observations of eight comets: C/2011 L4 Panstarrs, C/2012 S1 ISON, C/1996 Q1 Tabur, C/1999 S4 LINEAR, C/2002 O4 Hönig, C/2010 X1 Elenin, C/2012 T5 Bressi and C/1973 E1 Kohoutek.

All eight SLCs presented in this work are new and have not been previously published.

Except when it is otherwise stated, in this work we adopted the *envelope of the data set* as the correct interpretation of the observed brightness. There are many physical effects that affect comet observations such as twilight, moon light, haze, cirrus clouds, dirty optics, lack of dark adaptation, excess magnification, and in the case of CCDs, sky background too bright, insufficient time exposure, insufficient CCD aperture correction, and too large a scale. All these factors diminish the captured photons coming from the comet, and the observer makes an error downward, toward fainter magnitudes. There are no corresponding physical effects that can increase the perceived brightness of a comet. Thus the *envelope* is the correct interpretation of the data. In fact the envelope is rather sharp, while the anti-envelope is diffuse and irregular.

The envelope represents an ideal observer, using an ideal telescope and detector, in an ideal atmosphere.

A key to the photometric parameters measured in the SLCs is given in Paper II, and a short description is given next. (1) The magnitude at $\Delta$, R, $\alpha$, is denoted by $m_{COLOR}(\Delta, R, \alpha)$, where $\Delta$ is the comet-Earth distance, R is the Sun-comet distance, $\alpha$ is the phase angle and $\beta$ the phase coefficient:

$$m_{COLOR}(\Delta, R, \alpha) = m_{COLOR}(1,1,0) + 5 \text{ Log } \Delta + 2.5 \text{ n Log } R + \beta \alpha \qquad (1)$$

(2) q, the perihelion distance, is given in AU. (3) Q, the aphelion distance, also in AU. (4) The turn on point $R_{ON}$. (5) The turn off point $R_{OFF}$. (6) The sum of these two values $R_{DIFF} = +R_{OFF} - R_{ON}$. (7) The absolute magnitude before perihelion $m_V(1,-1)$. (8) The absolute nuclear magnitude $m_{V-NUC}(1,1,0)$

$$m_{V-NUC}(1,1,0) = m_{V-NUC}(\Delta,R,\alpha) – 5 \text{ Log } \Delta R – \beta \alpha \qquad (2)$$

(09) The Amplitude of the secular light curve

$$A_{SEC}(1,-1) = m_{V-NUC}(1,-1,0) – m_V(1,-1,0) \qquad (3)$$

$A_{SEC}$ measures the difference between the nuclear absolute magnitude and the total absolute magnitude. The minus sign in R, indicates observations pre-perihelion. $A_{SEC}(1,-1)$ is a measure of activity of a comet, and thus a proxy for age. $A_{SEC}(1,-1)$ is measured pre-perihelion to avoid the thermal wave effect after perihelion, that would have introduced thermal parameters for which we lack information. $m_{V-NUC}(1,1,0)$ is reduced to zero phase angle. (10) The diameter D. (11) The photometric age P-AGE(-1,1) in comet years (cy), defined below. (12) In the top line of the plot, right hand side V.year is the version of the plot and Epoch indicates the year of perihelion.

In this work we will be concerned only with the following parameters: q, Q, $R_{ON}$, $R_{OFF}$, $m_{V\text{-}TOT}(1,-1)$, $m_{V\text{-}NUC}(1,1,0)$ and $A_{SEC}(1,-1)$, that will be used to calculate the photometric age, P-AGE, the diameter D, and $T_{ON}$, $T_{OFF}$, needed as limits to integrate the water production rate to get the water budget (Section 7).

In the x-axis of the plots all logs are positive. Negative logs indicate observations pre-perihelion not values less than one.

In this work calendar dates will be expressed as YYYYMMDD. Although not continuous, this is a monotonically increasing number.

## 2.2 Photometric Age, P-AGE

The photometric age defined in Paper II is an attempt to define the age of a comet using activity as a proxy:

$$\text{P-AGE}(1,-1) = 1440 / [\, A_{SEC}(1,-1) \times R_{SUM} \,] \text{ comet years (cy)} \qquad (4)$$

P-AGE is measured in comet years that should not be confused with calendar years. The constant is chosen so that comet 28P/Neujmin 1 has a P-AGE = 100 cy.

## 2.3 Calculation of the Diameter, D

The absolute nuclear magnitude in the visual, $V_{NUC}(1,1,0)$, is related to the diameter D in km, by a compact and friendly formula derived in Paper II:

$$\text{Log}\,[\, p_V D^2 / 4 \,] = 5.654 - 0.4\, V_{NUC}(1,1,0) \qquad (5)$$

where $p_V$ is the geometric albedo in the visual. For comets for which the geometric albedo has not been measured, it is common to adopt $p_V = 0.04$. Thus the previous equation can be simplified even further:

$$\text{Log}\,[\, D^2 \,] = 7.654 - 0.4\, V_{NUC}(1,1,0) \qquad (6)$$

which is easy to remember.

### 2.4 A Lower Limit to the Nuclear Diameter

In Figure 1 we show $A_{SEC}$ vs Diameter for 29 comets (Ferrín et al. 2012). It is apparent that there are five comets with a maximum value of $A_{SEC}$, setting an upper limit $A_{SEC}$(Limit)=11.6±0.1. A practical consequence is that we can set a lower limit to the diameter of any comet if the absolute magnitude, $m_V$ (-1,1), is known, as is the case for most comets. Thus Equation (3) can be rewritten

$$11.6\pm0.1 > m_{V\text{-}NUC}(1,-1,0) - m_V(1,-1) \qquad (7)$$

Once $m_{V\text{-}NUC}$ (1,-1,0) has been determined from Equation (7) application of Equation (6) gives the lower limit to D.

## 3. Comet C/2011 L4 PANSTARRS

### 3.1 SLC

Figure 2 shows that C/2011 L4 Panstarrs turned on much before comet Halley (at $R_{ON}$= - 6.15±0.19 AU, Paper I). Since water cannot sublimate at distances beyond - 6 AU, the initial activity of the comet must be driven by an ice more volatile than water, probably CO or CO2. Alternatively the activity could be due to the amorphous to crystalline ice transition (Prialnik and Bar-Nun, 1987). However we have not been able to find evidence of this activity in 29 SLCs published in *The Atlas*.

Then, at R(SDE)= -4.97±0.03 AU, which corresponds to 20120414±3 d, the comet experienced a SDE. For comparison for 1P/Halley R(SDE)= -1.7±0.1 AU. Before the SDE, the power law was $R^{+8.67}$ and after the SDE it was $R^{+2.24}$ (Table 3).

We have previously shown (Ferrín, 2013b and Figure 1) that there is a maximum value to $A_{SEC}$, $A_{SEC}$(Limit) = 11.6±0.1. Since we know $m_V$ (-1,1) from the visual SLC in Figure 3, $m_V$(-1,1)= 5.6±0.1, we can calculate $m_{V\text{-}NUC}$ (1,1,0)= 17.2±0.2 using Equation 5. With a geometric albedo $p_V$ = 0.04 and Equation 3, we get a lower limit to the diameter D > 2.4±0.3 km.

To calculate P-AGE we assume that the turn off point is equal to the turn on point. Then we find P = AGE > 2.8 cy which indicates that this is a young comet. Combining this information with 1/a(original) = 0.000030 (Table 3), we conclude that this is a fresh, dynamically new, active comet coming from the Oort Cloud, and thus it is reasonable to assume that it is sublimating from 100% of its surface area (in reality 50%, the sunlit area). We will use this information to plot the comet in the RR versus ML-AGE diagram, but since the diagram is logarithmic, it is very forgiving. We could have a diameter twice the value, that the location in Figure 25 will not change by much.

In Figure 3, we see the difference $m_{MPCOBS}$ versus $m_{VISUAL}$. The plot shows that caution has to be exercised when using this data, since the MPCOBS and visual observations differ by a large amount. For example, it is the absolute magnitude from *visual* data that has to be used when calibrating the water production rate, not the magnitude from the MPCOBS data. Also, it is the *visual* absolute magnitude that has to be used to calculate a lower limit to the diameter in Equation 3. The observed difference $m_{MPCOBS}$ - $m_V$ ~ 1.2 to 3.4 mag. The comet abandons the power law and exhibits a brightness surge, near perihelion. The comet passes the $m_V$ (1,q) = 0 line and reaches to $m_V$(1,q) = -1.2±0.2, categorizing it as a *Great Comet* (those with negative $m_V$ (1,q)).

Figure 4 compares visual and CCD data. Observations still show a significant difference in magnitudes. This data also shows the SDE, but due to scatter, it is not possible to derive a precise date for the event.

Concerning the current status on cometary photometry, there is a technical problem that has not been yet solved, and that is the lack of agreement between CCD and visual data. Many observers extract fluxes with small apertures and then calculate dust production rates using the A'Hearn et al. (1984) formalism. As can be ascertained from Figures 2 to 4 and 6 to 9, these produce magnitudes too faint that do not reach to the envelope. Consequently all dust production rates measured with small apertures are actually lower limits. To solve this problem, we have proposed that a curve of growth method be used to produces infinite aperture magnitudes (Ferrín 2005b and paper II). The same problem happens with water production observations (Figure 18).

**3.2 Temperature**

Woodward et al. (2013) measured an infrared temperature of 312 K at R= -0.84 AU for comet C/2012 L4 Panstarrs while Sitko et al. (2013) measured the infrared temperature of comet ISON at R= -0.69 AU. Both temperatures are plotted in Figure 16 along with the temperatures of other comets, and both temperatures lie below the mean line. This could be due to the presence of larger particles or larger albedos than usual. We find T(All Comets)= 323±4ºK/SQRT(R), T(Panstarrs)= 287±4ºK/SQRT(R), T(ISON)= 273±4º K/SQRT(R).

**4.0 Comet C/2012 S1 ISON**

**4.1 Color**

Lisse et al. (2013) made a series of measurements in different bands. They found $m_V$ = 15.9±0.1, $m_R$ = 15.5±0.1, $m_I$ = 15.2±0.1 on June 11.16, 2013, which allows the determination of colors $m_V$ - $m_R$ = 0.4±0.14, and $m_R$ – $m_I$ = 0.3±0.14. The location of comet ISON in a color-color diagram is shown in Figure 5. The colors of ISON are consistent with the colors of other comets.

**4.2 SLC**

Figures 6 to 9 show the unusual SLC of comet ISON. To see how strange it is, compare with the SLC of comet C/1973 E1 Kohoutek, the famous comet that was erroneously said to have fizzled, which actually did not, shown in Figure 15. This is an entirely normal SLC typical of an Oort Cloud comet as confirmed in The Atlas I. The SDE+dip signature is clearly seen in all the Figures 6 to 9. The near-standstill of the comet after the SDE can only imply that the nucleus is depleted in CO or CO2. Table 1 compiles some statistics derived from the different data sets.

**4.3 Onset of Disintegration**

Using Figure 9, the visual secular light curve, it is possible to determine the onset of disintegration of comet ISON, marked by time t1 on the plot, when the comet increased its brightness by a factor of 22x. We find $T_{ONSET}$ = 13.25± 0.10 UT Nov. 2013, R(Dis) = -0.66±0.01 AU pre-perihelion. The same information is contained in Figure 22, the water production rate measure by Combi et al. (2013 b). We obtain $T_{ONSET}$ = 13.18± 0.10 UT Nov. 2013, at R= -0.68±0.01 AU. The agreement between the two data sets is excellent.

## 5. SLCs Of Comets C/1996 Q1 Tabur, C/1999 S4 LINEAR, C/2002 O4 Hönig, C/2010 X1 Elenin, C/2012 T5 Bressi and C/1973 E1 Kohoutek

The SLCs of these comets are shown in Figure 10 (C/2002 O4 Hönig), Figure 11 (C/1996 Q1 Tabur), Figure 12 (C/2010 X1 Elenin), Figure 13 (C/2012 T5 Bressi). All of them show the SDE+dip signature and all of them disintegrated suggesting that comet ISON will do so too.

Figure 15 shows the SLC of comet C/1973 E1 Kohoutek, the famous comet that was erroneously said to have fizzled. This is a misname. After the SDE the comet continued brightening at a significant rate, exhibiting a brightness surge at perihelion, reaching to $m_V$ (1,q) = -2.0, and thus into the *Great Comet Category* (comets with a negative magnitude at perihelion). So it did not really fizzle.

## 6. Why Should There Be a SDE?

When a comet from the Oort Cloud (e = 1.0) falls toward the Sun, the upper layer of the nucleus contains fresh volatiles like CO, CO2 and H2O. As the comet approaches, temperature increases, and the first one to sublimate is CO. Next sublimates CO2 and finally H2O. This is due to their different vapor pressures. When CO or CO2 is sublimating, the light curve far from the Sun is a straight line with a steep power law ~ $R^{+9.1\pm2.0}$ (Table 2). The H2O does not have sufficient temperature to sublimate and thus CO or CO2 controls the surface sublimation and the light curve. The sublimation rate increases as the comet approaches the Sun, and at a given temperature (near R ~ -2.8 AU), H2O overpowers CO or CO2, and H2O now controls the surface sublimation. This can be very clearly seen in the SLC of comet 9P/Tempel 1 presented in the Atlas I. The brightness increase decreases its rate according to the new sublimation rules. This is a plausible mechanism behind the SDE.

The m(SDE) versus R(SDE) Diagram is shown in Figure 17 and it is based on the data presented in Table 2. It shows 14 comets closely located in a narrow interval of this phase space 1.20 < R(SDE) < 2.12 AU, and five comets beyond this interval. Most of these comets are members of the Jupiter family but notice that 5 members of the Oort Cloud are located inside the Jupiter Family Interval. The Interval may represent the changeover from CO2 to H2O controlling volatile of the surface sublimation. It is not yet clear if the other five comets represent the CO to CO2 change over. Alternatively the changeover could be due to the amorphous to crystalline ice transition (Prialnik and Bar-Nun, 1987).

## 7. MASS-LOSS Budget and Mass-Loss AGE

### 7.1 Introduction

One way to assign an age to a comet is using the parameter $A_{SEC}$. $A_{SEC}$ diminishes as the comet gets older. Another way to assign an age is using the amount of mass loss by the object per orbit, as a proxy for age. It is to be expected that older objects are less active than younger objects. This mass loss (ML) is composed of water, CO, CO2 and dust. The objects may come from three repositories: the Oort Cloud, the Jupiter Family, and the asteroidal belt. However all produce gas and dust in large amounts. To calculate the water budget, WB, we will make use of water production rates Q, measured with a variety of techniques, from ultraviolet to radio.

A number of factors conspire to lower the measured water flux. For example a half power beam width smaller than the water coma size, or insufficient integration time, or insufficient CCD aperture error, or a small instrument. Combi et al. (2013 b) report water production rate measurements versus aperture size for comet C/2009 P1 Garradd at 2 AU from the Sun. The plot (Figure 18), shows an increase vs aperture with an asymptotic value. This is one reasons why it is advisable to adopt the envelope of the water production rate measurements, as the correct interpretation of the data. Something similar happened for photometric observations.

We define the Mass Loss Budget, ML-Budget, as the total mass expelled by the comet in a single orbit.

The Mass Loss Budget in kg, ML-Budget, is given by sum of the daily production rate values, from $T_{ON}$ to $T_{OFF}$ :

$$\text{ML-Budget} = \sum_{T_{ON}}^{T_{OFF}} Q_{GAS+DUST}(t) \,.\, \Delta t \qquad (8)$$

The sum goes from $T_{ON}$ to $T_{OFF}$ and this information is taken from the SLC plots. Let us define an age, the ML-Budget Age, ML-AGE, thus

$$\text{ML-AGE [cy]} = 3.58\text{x}10^{+11} \text{ kg / ML-Budget} \qquad (9)$$

In Equation (9) we have chosen the constant so that comet 28P/Neujmin 1 has a ML-AGE=100 cy. The ML-Budget and the ML-AGE is calculated for 29 comets in Table 4.

**7.2 Mass Loss, ML, and Remaining Returns, RR**

We have seen how to determine the comet mass loss per apparition in kg. However we are interested in the total mass loss. To calculate it we need the dust to gas mass ratio, δ.

Using a model, de Almeida et al. (2009) have derived production rates of gas and dust for several comets. Their results for comets 1P, 46P, 67P, and C/1996 B2, are particularly relevant to this investigation. Figure 19 shows that the dust to gas mass ratio is constrained to $0.1 < \delta < 1.0$, and that $\delta = 0.5$ is a mean value that fits the general distribution quite well, over a range of several orders of magnitudes. Thus we will adopt $\delta = 0.5$ for comet C/2011 L4 Panstarrs and other comets in Table 4. However the RR versus ML-AGE diagram (Figure 25) will show that the location of a comet on the diagram is not sensitive to the δ value. To get to the total Mass Loss Budget, we have to add the budgets of CO, CO2 and other volatiles that appear in smaller amounts.

With this information it is possible to calculate the thickness of the layer lost per apparition using the formula

$$\Delta r = ( \delta + 1 ) \text{ ML-Budget} / 4 \pi r_{NUC}^2 \rho \qquad (10)$$

where $r_{NUC}$ is the radius of the nucleus and $\rho$ its density. This equation can be derived from the density, given by $\rho = \Delta M/\Delta V$, the volume removed given by $\Delta V = 4\pi r_{NUC}^2 \Delta r$, and ΔM, the mass removed given ML-Budget. For the density we are going to take a value of 530 $kg/m^3$ which is the mean of 21 determinations compiled in Paper I. Then

$$RR = r_{NUC}/\Delta r \qquad (11)$$

The resulting values of Δr and RR are compiled in Table 4 for 29 comets. For example we see that comet 45P lost 9.7 m in radius per return. Since the radius of this comet is only 430 m, the ratio $r_{NUC}/\Delta r = 46$. This calculation implies that the comet will sublimate away in only 46 additional returns, if the mass loss rate continues at the present rate.

**7.3 Comet C/2011 L4 Panstarrs**

Figure 20 shows the water calibration of this comet, used to calculate the water budget. There are only two water measurements available in the literature by Biver et al. (2012) and Opitom et al. (2013). They coincide quite well with the Jorda, Crovisier & Green (2008) calibration, if the line is displaced downward by a factor of approximately 6.2x. Using this information a water b7udget is calculated in Table 4, and adopting D = 2.4 km as a first approximation to the diameter from Section 3, it is possible to plot the position of this comet in Figure 25. The object lies in the Oort Cloud region of the diagram.

**7.4 Comet C/2012 S1 ISON**

There are many measurements of the water production rate: Schleicher (2013), Combi et al. (2013 a), Weaver et al. (2013), Dello Russo et al. (2013), Opitom et al. (2013), Crovisier et al. (2013), Bodewits et al. (2013 a, b), Bonev et at. (2013), Keane et al. (2013) and Mumma et al. (2013). Transforming Figure 21 to a time plot and integrating, we find a Water Budget WB = $3.94 \times 10^{10}$ kg.

For comet C/2012 S1 ISON we will integrate the dust production rate compiled in Figure 24, extending the curve into perihelion using the visual light curve (Figure 9). Near perihelion the gas was exhausted and the brightness was only due to the dust. To convert from cm to kg we use the conversion factor 1000 cm = 1000 kg (A'Hearn et al., 1995). Several groups measured the dust production rate: Lisse et al. (2013), Opitom et al. (2013), Bodewits et al. (2013 a, b), Scarmato (2013), Fitzsimmons et al. (2013). In this way we found a dust budget DB = $2.35 \times 10^{11}$ kg.

For the CO production rate we use the data shown in Figure 23. The result is CO-Budget = $2.5 \times 10^{9}$ kg. There is no firm evidence of the existence of $CO_2$.

### 7.5 Diameter of Comet ISON

Adding all the contributions plus a 10% of gas from other lesser volatiles, we find a total Mass Loss Budget ML-Budget = $3.05 \times 10^{11}$ kg. If we assume a density of 530 kg/m$^3$ which is the mean value of 21 determinations (Ferrín, 2006) it is possible to determine a diameter. We find D = 1030±70 m. This diameter agrees very well with the upper limit found by Delamere et al. (2013), D(mean) < 1126 m.

Using Equation 6 this diameter implies an absolute magnitude $m_{V\text{-}NUC}$ (1,-1,0) < 19.1. In our Figure 9 it is apparent that the absolute magnitude of comet ISON was $m_V$(1,-1) = 8.1±0.1. Using Equation 3 we then find $A_{SEC}$ (1,-1) > 11.0 which allows plotting this comet in the $A_{SEC}$ vs D diagram, Figure 1.

With the above information we also find a dust/gas ratio $\delta = 6$. This mean value *for the whole apparition* is plotted in Figure 19 and implies that this comet was very dusty.

Using Equation 9 we find a Mass Loss Age ML-AGE = 1.2 cy. Using Equation 4 and Figure 9, the photometric age can be calculated and we find P-AGE(1,-1) = 0.32 cy, a exceedingly young comet, surpassing comet Hale-Bopp (P-AGE= 2.3 cy) and thus holding the record of youngest comet in our database (see the Atlas I). Both methods agree and this result is also in agreement with the inverse semi-major axis of the orbit 1/a = 0.000009 that suggests a dynamically new comet.

### 7.6 Comet C/2002 O4 Hönig

This is a dynamically new Oort Cloud comet as can be deduced from the inverse semi-major axis of the orbit, 1/a = -0.000772. In fact this is the most hyperbolic comet of the list, what prompts us to question if the extraordinary nature of this object, may be related to its 1/a value.

The disintegration of this object was registered by many observers and was analyzed in detail by Sekanina (2002), who concluded that the total amount of dust expelled was 1 to $2x10^{10}$ kg. For our work we will adopt 1.5 $10^{10}$ kg of dust. He also assigns a probable diameter of 0.7 km, if the object is made of CO.

If we adopt $\delta = 1$, then the mass of gas was identical to the mass of the dust and ML-Budget = $3x10^{10}$ kg, ML-AGE = 24 cy, and RR = 13. However if we adopt $\delta = 0.1$, then the mass of gas is 10 times the mass of dust, and we get ML-Budget = 1.65 $10^{11}$ kg, ML-AGE = 2.4 cy and RR = 1.7. Since the comet disintegrated (Sekanina, 2002), obviously RR = 1 (Figures 25).

## 8. The Remaining Returns vs the Mass Loss Age Diagram

The Remaining Returns versus Mass Loss Age diagram (Figure 25) (Ferrín et al, 2012; 2013a), is an evolutionary diagram that makes use of the Mass Loss Budget (the total amount of gas and dust expelled by the comet per orbit), and the diameter of the comet. These two parameters have to be calculated in advance to plot a comet on the diagram.

Additional comets have been added to the original RR versus ML-AGE diagram (Ferrín et al. 2012; 2013a). These objects illustrate the complexity of the diagram. The complete description of the diagram has been moved to the Caption of Figure 25.

For a *sublimating away comet*, the thickness of the layer removed each apparition should remain constant as a function of time, as can be seen from the following argument. The energy captured from the Sun depends on the cross section of the nucleus, $\pi\ r_N^2$, on the Bond Albedo, $A_B$, and on the solar constant, S. The energy conservation equation can be written:

$$(1- A_B)\ S\ \pi\ r_N^2 = p_{IR}.\sigma.T^4 + K1.4.\pi.r_N^2 \cdot \Delta r_N . L + K2\ \partial T/\partial x$$

where $A_B$ is the Bond albedo, S the solar constant, $r_N$ the nuclear radius, $p_{IR}$ is the albedo in the infrared, T the temperature, K1 and K2 are constants, and σ the Stephan-Boltzmann constant, L the latent heat of sublimation, $\Delta r_N$ the thickness of the layer removed, x the depth below the surface.

The term on the left is the energy captured from the Sun. The first term on the right is the energy radiated the second the energy sublimated, and the third the energy conducted into the nucleus. The first and third terms on the right hand side are small in comparison with the second term because at large distances the temperature is very low. The second term dominates near the Sun at perihelion. So in first approximation

$$(1- A_B)\, S\, \pi\, r_N^2 \sim K1\, 4\, \pi\, r_N^2\, \Delta r_N\, L$$

$$\Delta r_N \sim (1- A_B)\, S / 4\, K3\, L \qquad (12)$$

We find that $\Delta r_N$ should be approximately a constant, assuming that the orbit does not change, that there is no change in the active fraction of the nucleus surface, and that the pole orientation remains stationary.

Notice that r/Δr would tend to zero as the comet sublimates away. However if the comet contained much dust, part of it would remain on the surface, Δr would tend to zero due to suffocation and r/Δr would tend to infinity. Thus, sublimating away comets tend to zero and suffocating comets tend to infinity (Figure 25). Sublimating away comets move down and suffocating comets move up on the diagram.

**9. Evolutionary Lines**

Comets move on the RR versus ML-AGE diagram. It is a work beyond the scope of this paper, to calculate trajectories on the plot. Models could show rather complicated behavior and nonlinear motion if for example the pole orientation is changed by jets, as has been seen to take place. Additionally, comets experience jumps in perihelion distance due to planetary perturbations that will show up in the trajectories.

However, it is possible to get a *preliminary* idea of what this motion might be, by considering a very simple model of a sublimating comet. The more complex problem of a suffocating comet is beyond the scope of this work.

In the previous Section 8 we have shown that in the case of a surface sublimating with no dust left on the surface, the surface layer removed should be a constant as a function of time. Then if we start with a given radius, it is possible to calculate RR and ML-AGE as a function of time, and plot the result in Figure 25. We found that in this simple model trajectories are straight lines with negative slope, isolines.

An isoline can be defined as an evolutionary line that assumes that the comet losses a constant $\Delta r$ at each return and eventually sublimates away.

Suffocating comets have positive slopes. If the slope is negative, then there will be a time when the evolutionary line will intersect the RR = 1 line, the disintegration line. The intersection is the Death Age, DA, in units of comet years.

In this fashion we find DA(Kohoutek) = 1.7E06 cy, DA(NEAT) = 1.4E05, DA(Panstarrs) = 36000cy, DA(Hönig) = 6 cy.

One puzzling aspect of this calculation is that comet Hönig reached its RR = 1 line at the early age of DA = 6 cy (Figure 25). The youngness of the comet is confirmed by its 1/a value, the largest of the whole group in Table 2, and by the fifth position in Table 4 of a total of 29 comet, in terms of production rate.

Sekanina (2002) favors a CO composition and an explosion. In view of the present investigation, another hypothesis comes forward. Comet Hönig may have been a very young (pristine) hyperbolic comet coming from the Oort Cloud, made mostly of CO ice, that was exhausted as a consequence of its approach to the Sun. This hypothesis explains a number of features, better than the former one.

### 9.1 The Suffocation-Sublimation Border, SSB

Since suffocating comets move upward in the RR versus ML-AGE diagram (Figure 25), and sublimating comets move downward, there must be an intermediate value where motion must be horizontal, a *suffocation-sublimation-border*. We estimate the border at RR(SSB) = (6±5) $10^4$ . Notice the large error. At the present moment the border is so wide, that we do not know on what side of the border comets Hale-Bopp or C/2009 P1 Garradd are located.

## 9.2 Conversion From Comet Years To Earth Years

It is easy to convert from comet years to Earth years. For example in the vertical axis the remaining returns of comet 1P/Halley are RR(1P) = 1158. We know the orbital period of this comet is 76 years. Thus, assuming the comet follows the isoline, the extinction date will be 88008 Earth's years. On the horizontal axis of Figure 25, the isoline intercepts the RR = 1 line at DA(1P) = $1.1 \times 10^6$ cy. Thus the conversion for comet 1P/Halley is 12.5 cy/y.

For comet C/1996 B2 Hyakutake, a = 951 AU, $P_{ORBITAL}$= 29300 y. From Figure 22, the vertical axis tells us that this comet has 272 returns left. That is $8.0 \times 10^6$ Earth's years. On the other hand the horizontal axis of Figure 25, tells us that the Death Age is DA(HY) = $1.1 \times 10^5$ cy. Thus for this comet the conversion is 72.5 cy/y.

Once again it must be emphasized, that these calculations are valid only if comets follow the isolines, which has not yet been demonstrated observationally. For example comet 2P/Encke does seem to be following an isoline, but comet 103P/Hartley 2, clearly is not (Figure 25).

Now we see the advantage of using comet years. Comet years are the same for all comets, while calendar years are not.

## 9.3 The Desert

Figures 25 and 26 also clarify the concept of *desert.* Observationally, we do not have any comet plotted in the lower right hand side of the diagram. We would expect sublimating comets to sublimate away in a time scale much shorter than for suffocating comets to suffocate.

Theoretically (Figure 25), the suffocation-sublimation border line, will set a remaining returns value, RR(SSB). This value in turn fixes an isoline, of several mega-comet years. This line intercepts the RR = 1 line at the Dead Age of that several mega-comet years. This is the lower limit of the desert. At the present moment the comet that sets the limit is 2P/Encke, with a death age DA = $3.9 \times 10^9$ cy, that is $3.9 \times 10^9$ *cy < Desert.*

Clearly, this evolutionary model is simple, incomplete, sketchy and imperfect. In fact it is a back of an envelope calculation. However it shows several virtues and that is an interesting accomplishment. Much theoretical and observational work remains to be done to elucidate the full meaning and the limitations of the RR versus ML-AGE diagram.

## 10. Conclusions

The scientific results found in this investigation are:

**(1)** We reduced 16.673 photometric observations of comets C/2011 L4 Panstarrs, C/2012 S1 ISON, C/1996 Q1 Tabur, C/1999 S4 LINEAR, C/2002 O4 Hönig, C/2010 X1 Elenin and C/2012 T5 Bressi, and C/1973 E1 Kohoutek, and we present their secular light curves (SLCs). The eight SLCs are new and have not been published previously. The first two comets turned on beyond -10 AU from the Sun. For comparison comet 1P/Halley turned on at R= -6.2±0.1 AU. Since water ice can not sublimate at distances R < -6 AU, these comets have to contain substances more volatile than water, like CO or CO2.

**(2)** We measured the *Slope Discontinuity Event* (SDE) of C/2011 L4 Panstarrs (Figure 2). This is the distance at which the brightness increase rate slows down to a more moderate pace and it is reminiscent of the same process for comet 1P/Halley and 11 other comets listed in Table 2. We find R(SDE) = -4.97±0.03 AU, or t(SDE)= 2012 04 11±3 d. For comet 1P/Halley R(SDE) = -1.7±0.1 AU.

**(3)** We derive the absolute magnitude of C/2011 L4 Panstarrs and the power laws that define its brightness behavior (Figures 2 to 4). The absolute magnitude is $m_V(1,-1)$= +5.6±0.1 compared with $m_V(1,-1)$= +3.7±0.1 for comet 1P/Halley. After passing the SDE, the comet is increasing its brightness with a shallow power law $R^{-2.24}$. The magnitude at perihelion can be measured from the SLC and we find $m_V(1,q)$= -1.2±0.2 giving it entrance to the Great Comet Category (those comets with negative magnitudes at perihelion). Additionally the comet exhibited a perihelion surge (Figures 2 to 4).

**(4)** We measured the SDE of C/2012 S1 ISON. We find R(SDE) = -4.09±0.06 AU (which corresponds to 2013 04 15±7 d) (Table 1). We also measure the absolute magnitude. We find $m_V(1,-1)$= +8.1±0.1.

**(5)** Technically speaking, since April 15±7 d, 2013, comet ISON was at a standstill in brightness for more than 132 d, a rather puzzling occurrence (Figures 6 to 9). We found five comets with similar behavior: C/1996 Q1 Tabur, C/1999 S4 LINEAR, C/2002 O4 Hönig, C/2010 X1 Elenin and C/2012 T5 Bressi, all of which disintegrated. Thus there is a significant probability that comet C/2012 S1 ISON may disintegrate at perihelion. The future of this comet does not look bright. Note: The comet disintegrated as predicted (CBET 3731), while this paper was being refereed.

**(6)** For comparison we present the SLC of comet C/1973 E1 Kohoutek, the famous comet that was erroneously said to have fizzled, and show reasons to conclude that it didn't.

**(7)** We compiled published production rates of water, dust and CO and used them to calculate the Mass Loss Budget. We use this information for ISON and other data to calculate the diameter adopting a density. We find D = 1.03±0.07 km in excellent agreement with the upper limit found by Delamere et al. (2013) who found D(mean) < 1.126 km.

**(8)** We also plot these two comets in an Evolutionary Diagram that separates comets by class (Figure 25). The comets lie in the RR versus Mass Loss Age diagram, in the region of the Oort Cloud comets (the left part of the diagram). Since this is a Log-Log plot, the diagram is forgiving, and the results are robust. This is a complex diagram that contains a lot of information.

**(9)** Figure 26 allows the determination of the Death Ages of several comets, assuming they will continue evolving along an isoline. We have measured DA(KO)=1.7E06 cy, DA(V1)=1.4E05, DA(L4)= 36000cy, DA(Hö)=6 cy.

**(10)** Since suffocating comets move upward in the RR versus ML-AGE diagram (Figure 25), and sublimating comets move downward, there must be an intermediate value where motion must be horizontal, a suffocation-sublimation-border. We estimate the border at RR(SB)=(6±5) $10^4$ . At the present moment the border is so wide, that for example, we do not know on what side of the border comets Hale-Bopp or C/2009 P1 Garradd are located.

**(11)** *The Desert of Comets* is the right hand, lower part of the diagram were we should expect to find few or no comets. We estimate a theoretical value for the start of this region at 3.9 $10^9$ cy < Desert.

**(12)** There are many questions to be answered that have to deal with the RR versus ML-AGE diagram. However, two questions may be representative: Is it possible to calculate an evolutionary model containing all important physical phenomena, capable of predicting the long term motion of comets in the diagram? In particular, why is comet 103P moving in that particular direction, $60^o$ to the isolines?

Additional comet data and results can be found at the following web page: http://astronomia.udea.edu.co/cometspage/

**Acknowledgements**

I thank the referee Paul Weissman for his many suggestions that improved the scientific quality of this manuscript enormously. The observations reduced in this work were taken by hundreds of observers seeking faint specs of diffuse light at odd hours of the night. We thank CODI of the University of Antioquia for their support through project E01592. To Julio Castellanos of the Cometas-Obs list, for his permission to use their data sets.

**Appendix A**

The data sets to create the secular light curves are available in the internet. These sites release their datasets freely to the public, in accord with good scientific practices.

(1) The Cometary Science Archive http://www.csc.eps.harvard.edu/index.html is maintained by Daniel Green (Green, 2013).

(2) Another useful site is the Minor Planet Center repository of astrometric observations, http://www.minorplanetcenter.net/db_search . Tim Spahr is the Director of the Center.

(3) Seiichi Yoshida's web place http://www.aerith.net/ contains many raw light curves and has access to oriental sites difficult to translate. His own observations can be found here: http://www.aerith.net/obs/comet.html#2012S1

(4) The Yahoo site https://groups.yahoo.com/neo/groups/CometObs/info contains up to date observations by many observers for many comets.

(5) The group from Spain measures magnitudes with several CCD apertures:

http://www.astrosurf.com/cometas-obs/ It is managed by Julio Castellanos, Esteban Reina and Ramon Naves.

(6) The site http://www.cobs.si/ contains many observations in the ICQ format (International Comet Quarterly), as well as news concerning comets, and it is maintained by the Crni Observatory with Jure Zakrajsek as curator.

(7) The German comet group publishes their observations at http://kometen.fg-vds.de/archive.htm The editor is Uwe Pilz.

(8) Observers from South America collect their observations here: http://rastreadoresdecometas.wordpress.com/ This is the web site of LIADA (Liga Ibero-Americana de Astronomía) managed by Luis Mansilla.

(9) The site http://www.shopplaza.nl/astro/cometobs.htm contains observations of many comets and it is administered by Reinder J. Bouma and Edwin van Dijk.

(10) The site http://www.rea-brasil.org/cometas/ is the repository of cometary observations by observers from Brazil.

(11) The site http://www.brucegary.net/ISON/ mantained by Bruce Gary, contains photometric information on comet ISON.

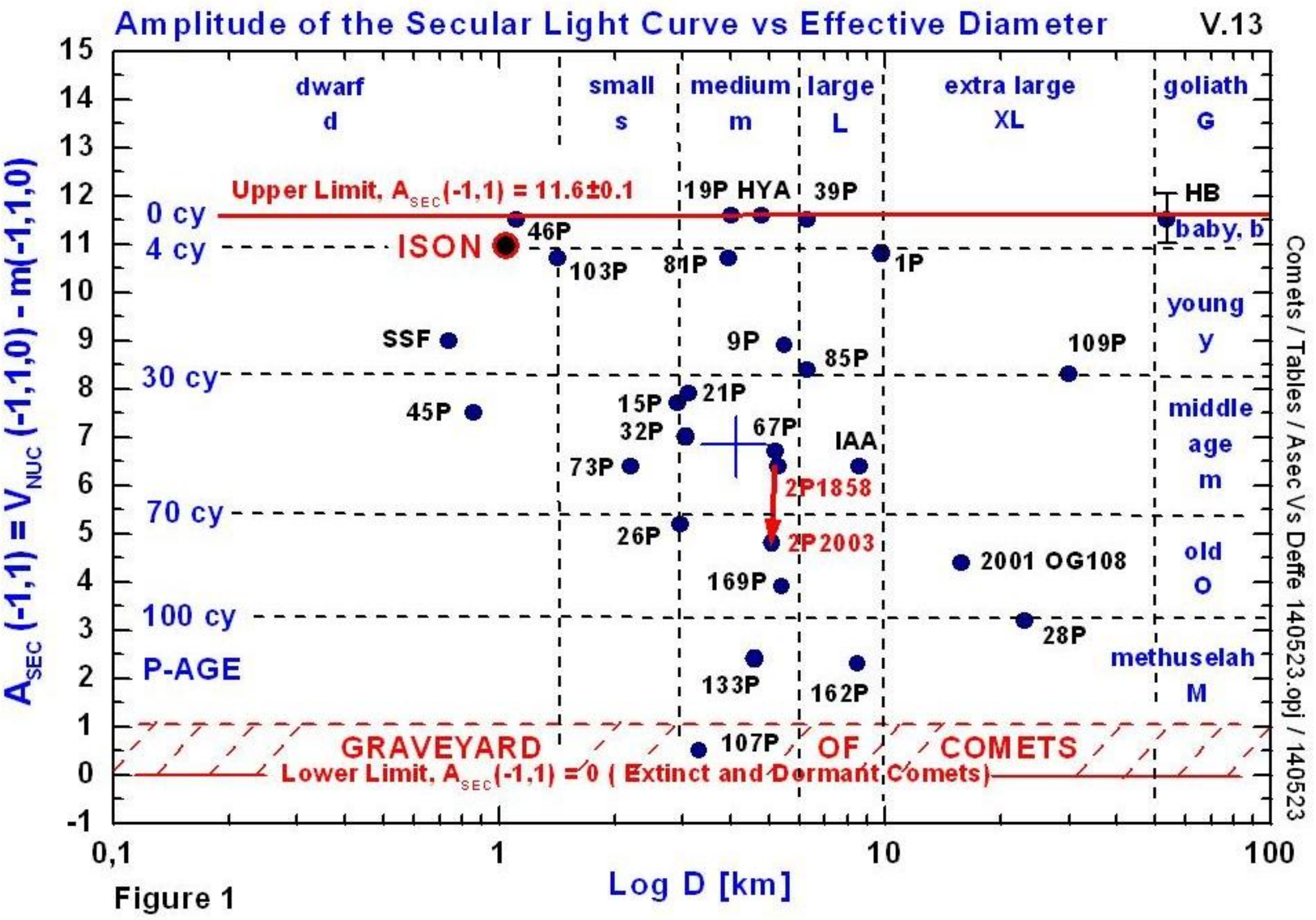

***Figure 1**. The $A_{SEC}$ vs Diameter relationship.* *Comets are placed in categories of diameter and photometric age, P-AGE, defined in Equation 3. Photometric Age increases downward. Five comets at the top show the same upper limit. In this work we calculate the Mass Loss Budget of comet ISON, $3.05 \times 10^{11}$kg, and assuming a density of 0.53±0.10 gm/cm$^3$ which is the mean value of 21 comets (Ferrín, 2006) we find a diameter D = 1.03±0.07 km. Then using Equation 6, $V_{NUC}$ = 19.1±0.1. In Figure 9 we find an absolute magnitude $m_V(1,-1)$ = 8.1±0.1, and using Equation 3 $A_{SEC}$ = 11.0±0.1. Thus the location of ISON in this plot can be found and it is displayed. The location near the upper limit implies that this is a baby comet sublimating from 100% of its surface area (actually 50%, the illuminated area). This Figure is updated from Ferrín et al. (2012).*

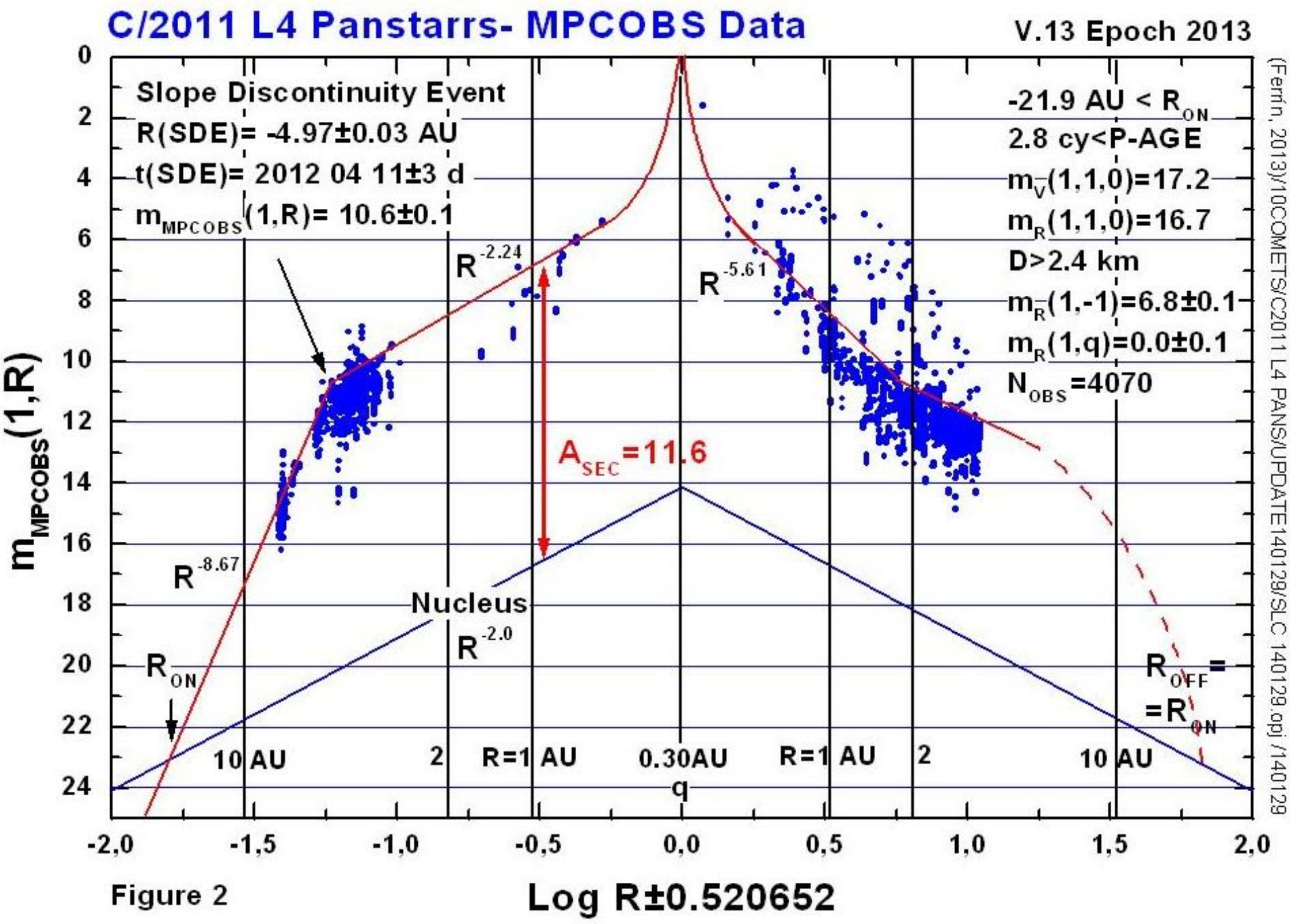

***Figure 2. The secular light curve (SLC) of comet C/2011 L4, log plot.*** *In this and the following plots the envelope of the data is used as the correct interpretation of the light curve (see text). The data shows a clear SDE at R(SDE)= -4.97±0.03 AU, which corresponds to 20120411±3 d. The slopes before and after SDE are measured. The nuclear line is calculated assuming $A_{SEC}$(Limit)= 11.6 (from Figure 1), and sets a lower limit to the diameter of the nucleus, D>2.4 km. Since the photometric age is small (P-AGE>2.8 cy), it is reasonable to assume that it is sublimating from 100% of its surface area. In this and the following plots in the x-axis all logs are positive. Negative logs indicate observations pre-perihelion not values less than one. The data set of this plot comes from the MPCOBS site.*

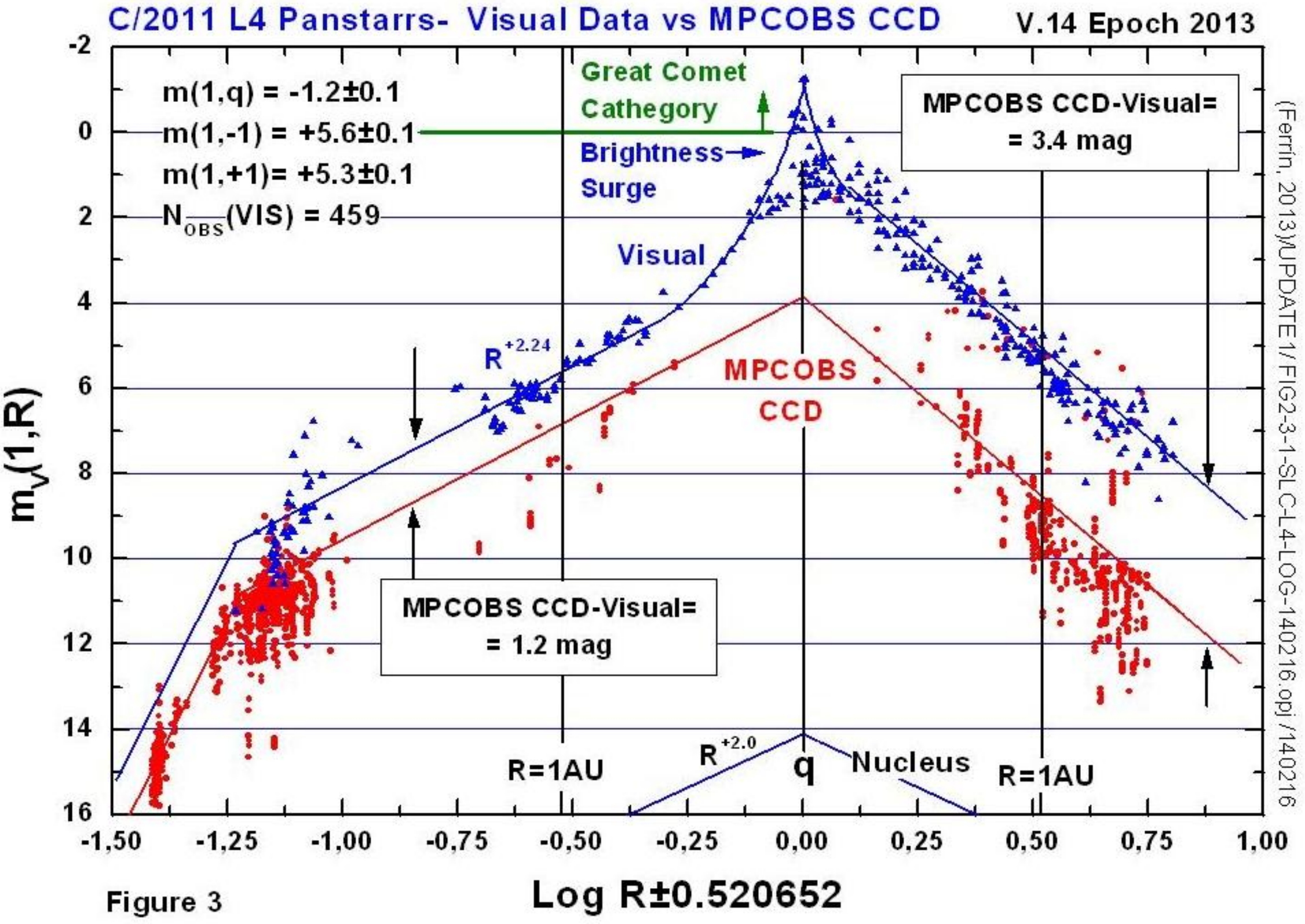

***Figure 3. SLC of C/2011 L4 Panstarrs, Log plot, visual data compared to CCD.*** *The plot shows that the difference $m_{VIS}$ - $m_{MPCOBS}$ is different before than after perihelion. This is clear evidence of the insufficient CCD aperture error (Paper II). The difference is larger after perihelion because the comet was larger in size and thus the whole flux could not be captured by CCD observations. Data bases like MPCOBS contain measurements that are a byproduct of astrometry. Photometric measurements are encapsulated in the software, using fixed and small apertures that do not capture the whole flux. The data set for this plot comes from the Minor Planet Center visual data, site (1) in the Appendix.*

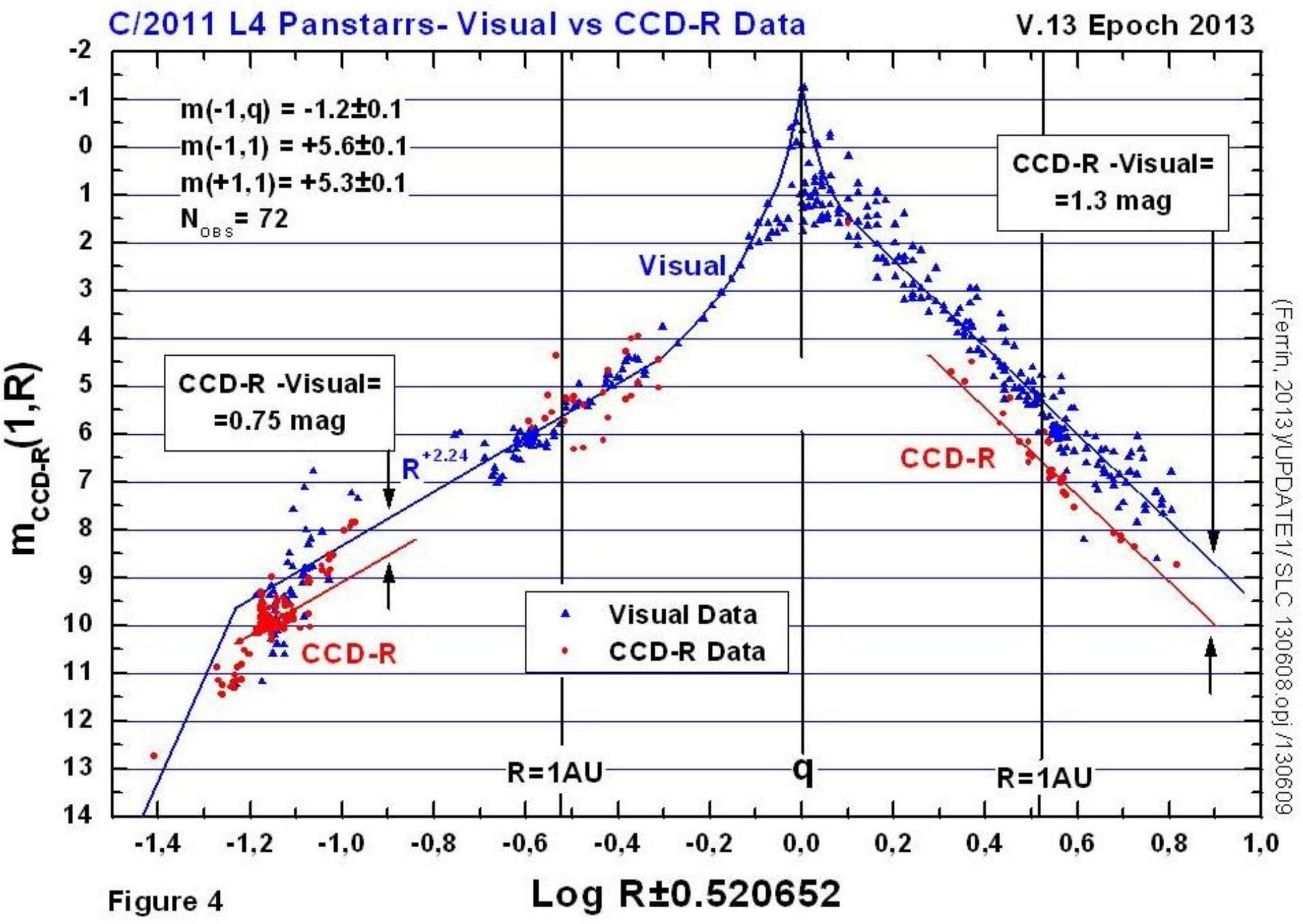

***Figure 4. SLC of C/2011 L4 Panstarrs, Log plot, Visual compared to CCD-R data.*** *CCD observations still show a significant difference with visual data. This data also shows the SDE, but due to scatter, it is not possible to derive a precise date for the event. The data for this plot comes from sites (4) to (9) in the Appendix.*

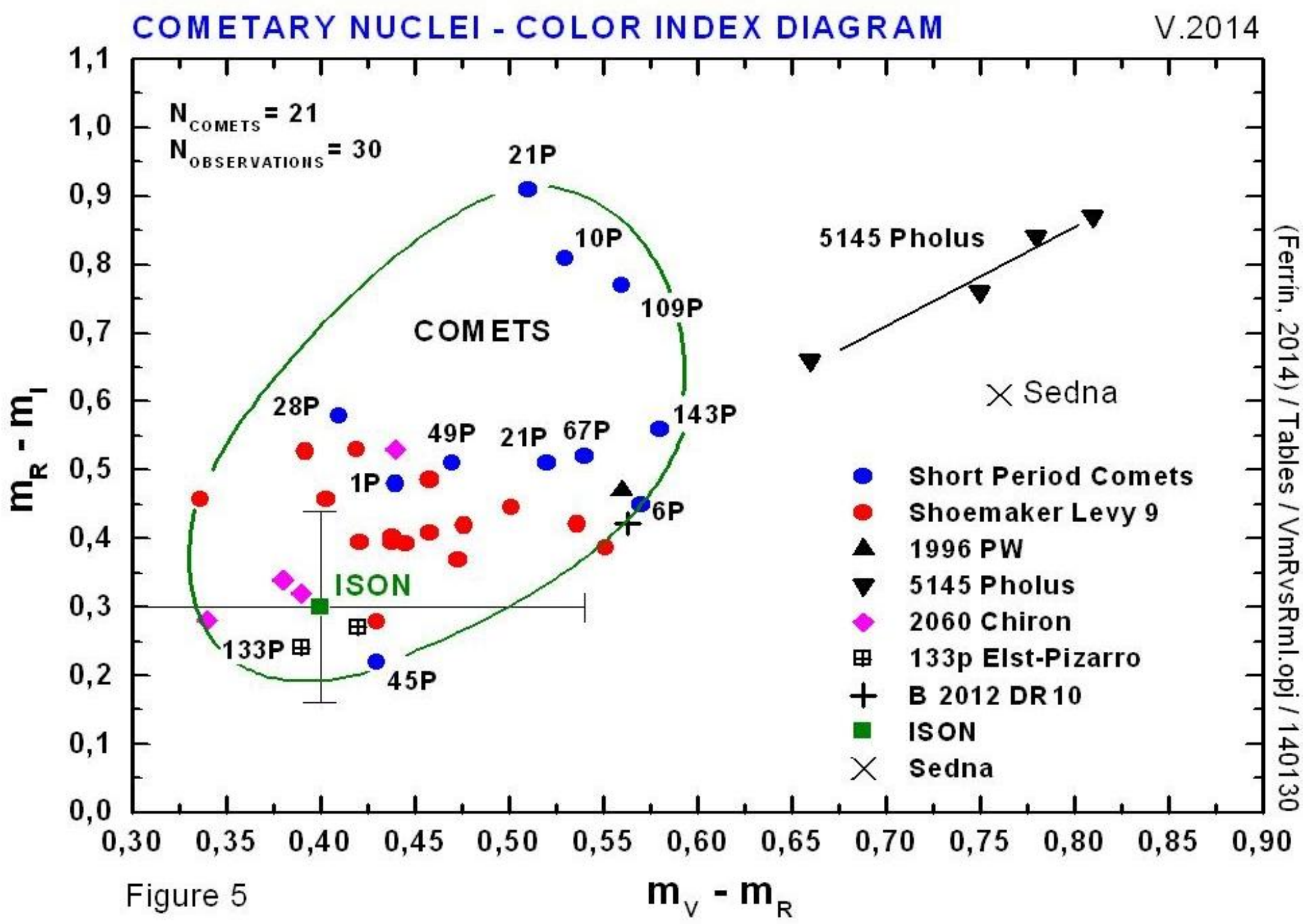

***Figure 5. Color index diagram for comets.*** *The location of comet C/2012 S1 ISON is shown and it is inside the area of localization of other comets. Cometary data compiled by Ferrín (2006). ISON's data from* Lisse et al. (2013)*.*

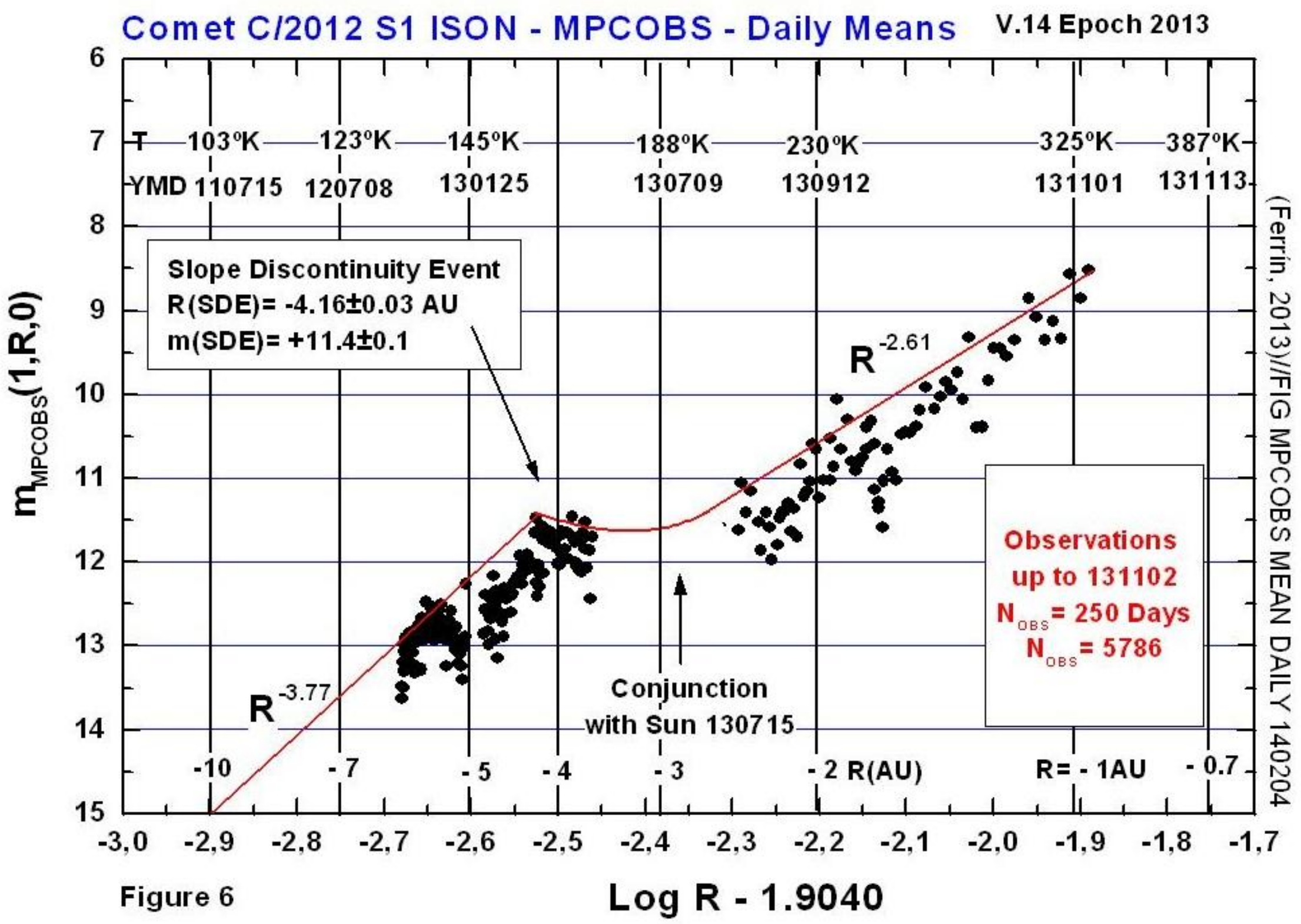

***Figure 6. The strange Secular Light Curve of comet C/2012 S1 ISON.*** *To see how strange this SLC is compare with the SLC of comet Kohoutek shown in Figure 15. To avoid the vertical dispersion exhibited by this dataset, we have taken daily mean values of the data. The envelope is down by +0.66 mag with respect to the non-averaged data. Pre-Conjunction with the Sun, the comet exhibited a SDE+near-standstill signature. The temperature above has been calculated from the formula T = 324ºK / SQRT(R) where R is the distance to the Sun in units of AU (Ferrín et al., 2012). Since these data are mean daily values and averages measurements from different filters, it is not being used for any scientific study other than determining the location of the SDE and its general shape. 5786 observations of the MPCOBS database were averaged daily (site 2 in the Appendix).*

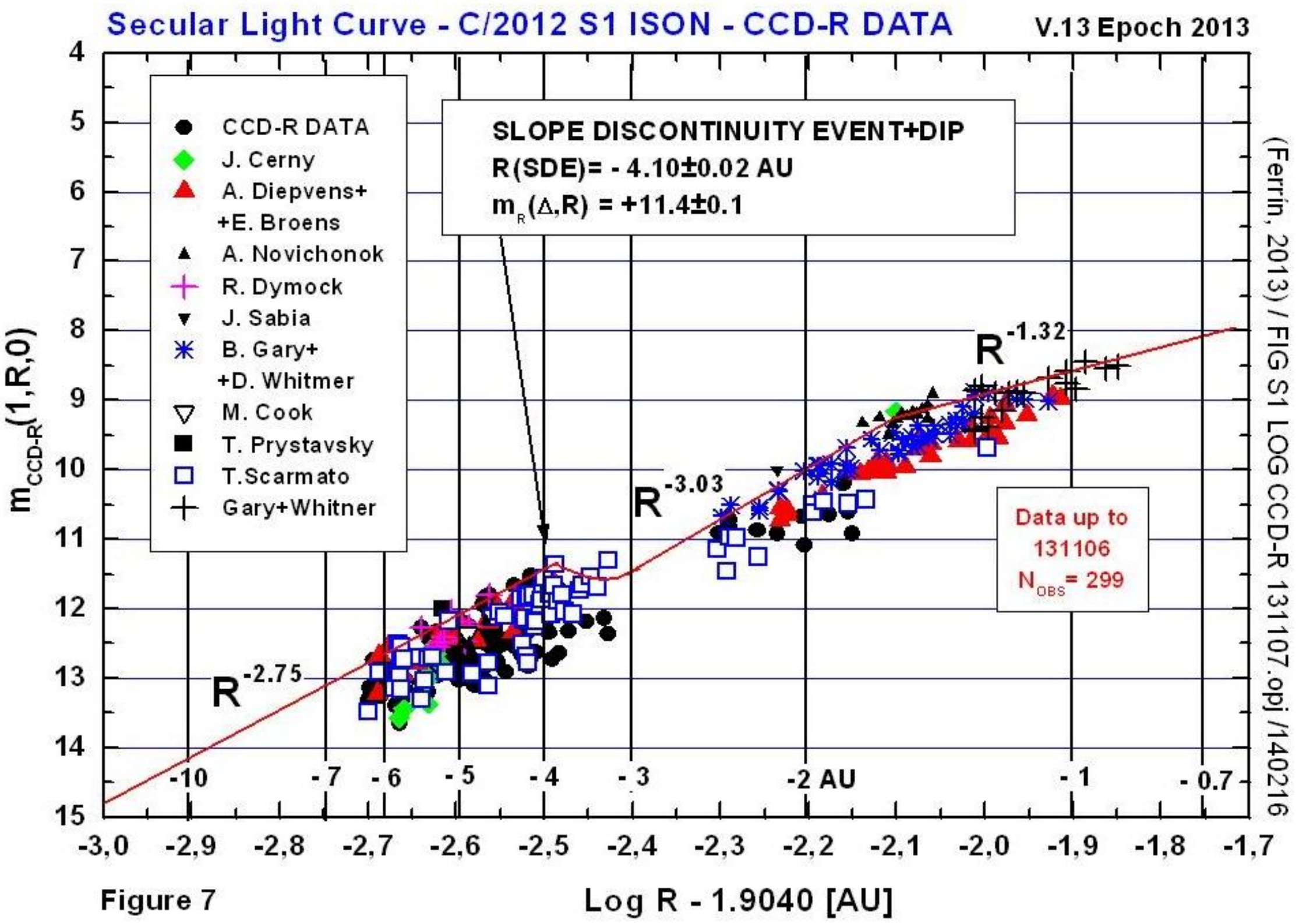

***Figure 7. The strange secular light curve of comet C/2012 S1 ISON, CCD-R data versus log of distance to the Sun.*** *The comet exhibits a Slope Discontinuity Event (SDE) plus a near-standstill in the light curve that is confirmed independently in other data sets. An absolute magnitude in the red band pass can be deduced from this plot. Any power law with power less than -2 implies that the comet is fading. In this dataset the comet is fading for a long period of time before perihelion. Since CCD-R measurements are of high accuracy this result is robust.*

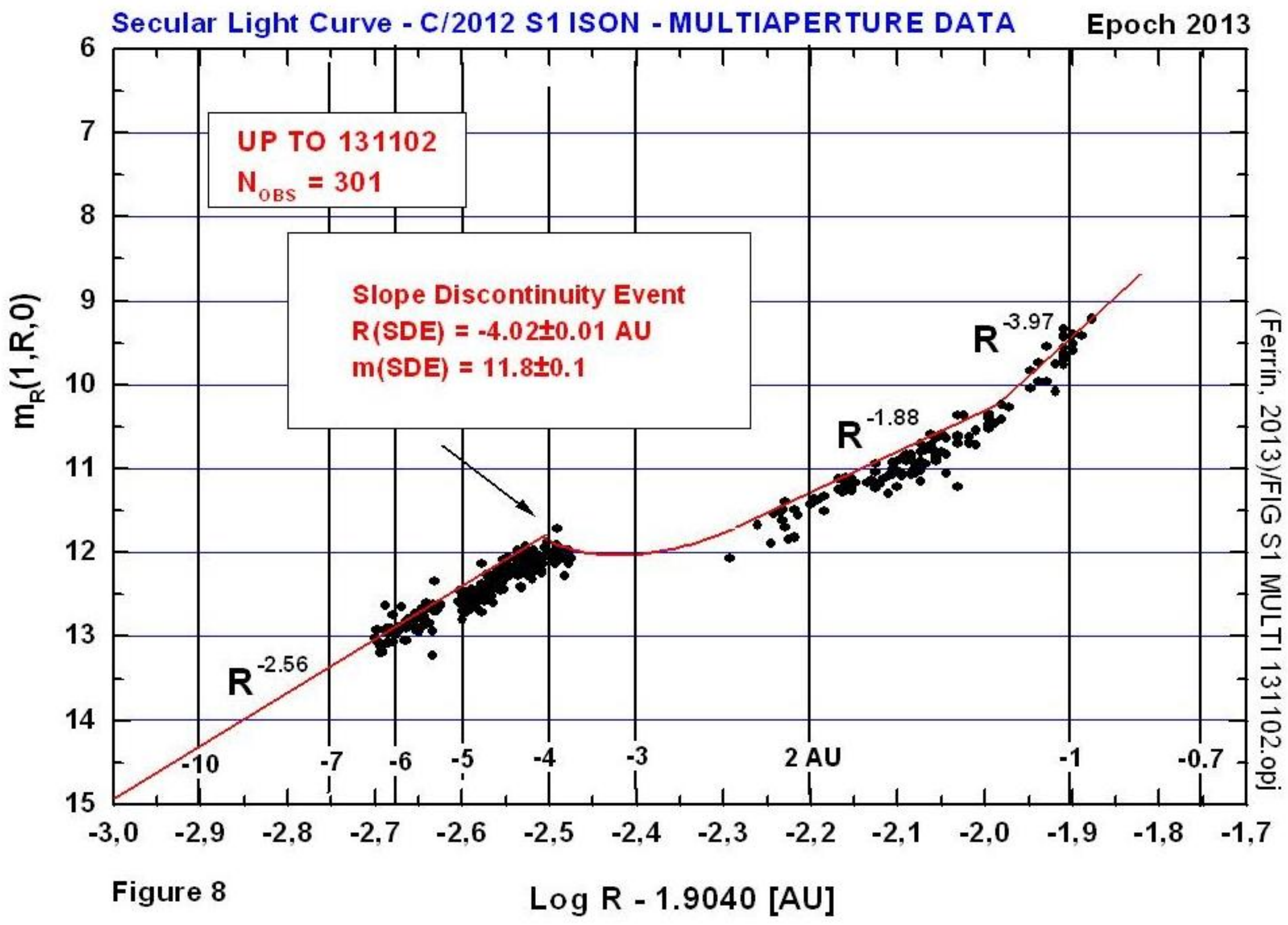

***Figure 8. Comet C/2012 S1 ISON, cometas-obs CCD data,*** **averaged daily.** *This data set obtained using mean daily values of multiple aperture observations, shows* independently *the SDE+ near-standstill signature shown in the previous Figures. Thus it must be real. Any power law with power less than -2 implies that the comet is fading. The data for this plot has been extracted from site (5 ) of the Appendix.*

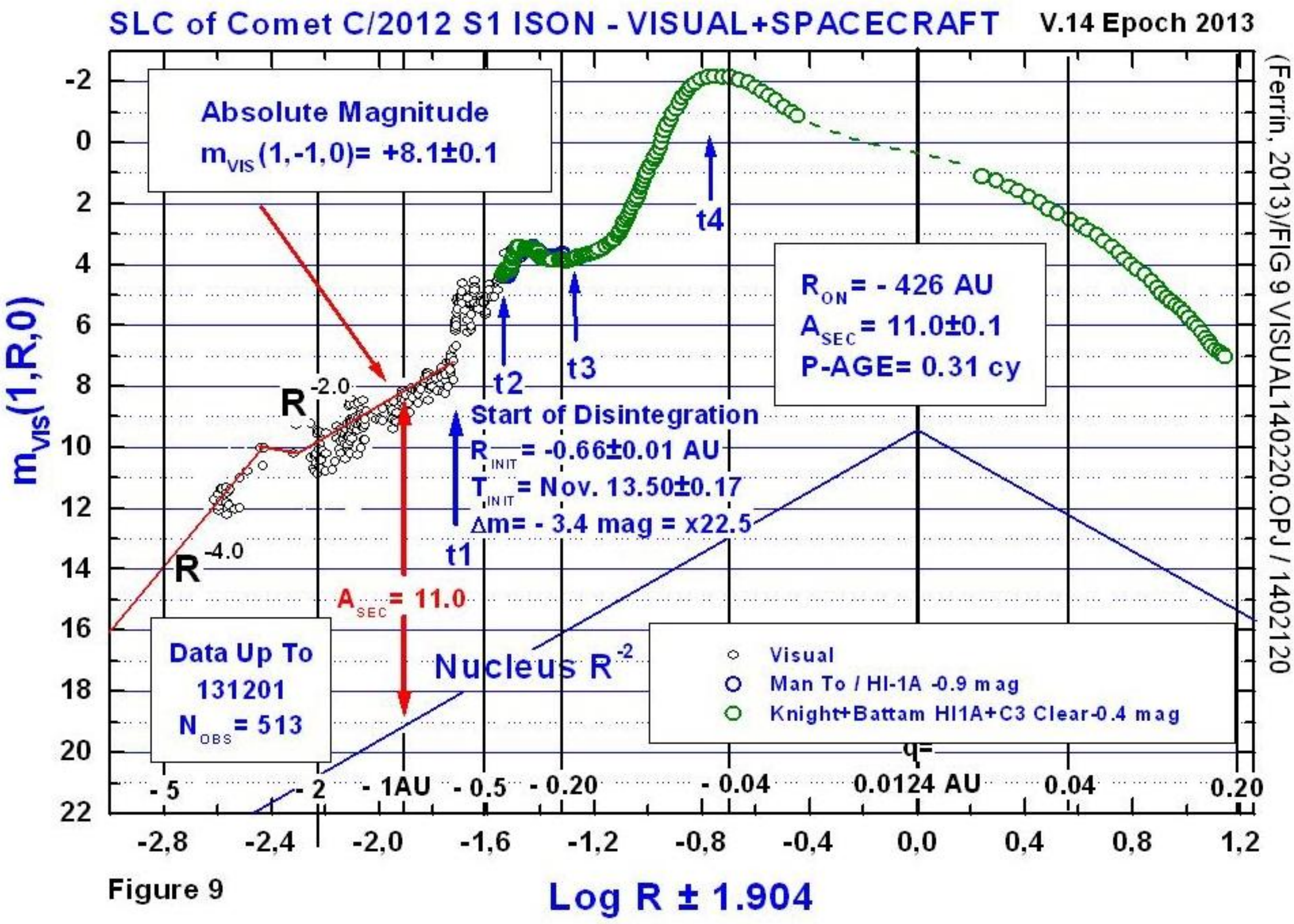

***Figure 9. The whole SLC of Comet C/2012 S1 ISON, visual data.*** *The absolute magnitude derived from this dataset is* $m_V(1,-1,0) = 8.1\pm0.1$. *The nucleus line is drawn from the diameter determined in this work D = 1.03±0.07 km. Any power law with power less than -2 implies that the comet is fading. The power law of the light curve just after discovery is* $R^{-4}$ *while the nucleus line is* $R^{-2}$ *. These two lines intercept at R = - 426 AU pre-perihelion. Although this number is large, it suggests that the only volatile that could be responsible for activity is CO. This plot allows the determination of the onset of disintegration (signaled by time t1 in the plot) when the comet increased its brightness by a factor of 22x. We find R(Dis) = -0.66±0.01 AU pre-perihelion. Time t2 marks a smaller fragmentation event, while time t3 signals the initiation of the expansion of the dust cloud. At time t4 the optical thickness of the dust reaches a maximum and from there on the solar radiation pressure disperses and dissipates the cloud. The data for this plot comes from sites (4) to (9) in the Appendix and from Knight and Battam (2014).*

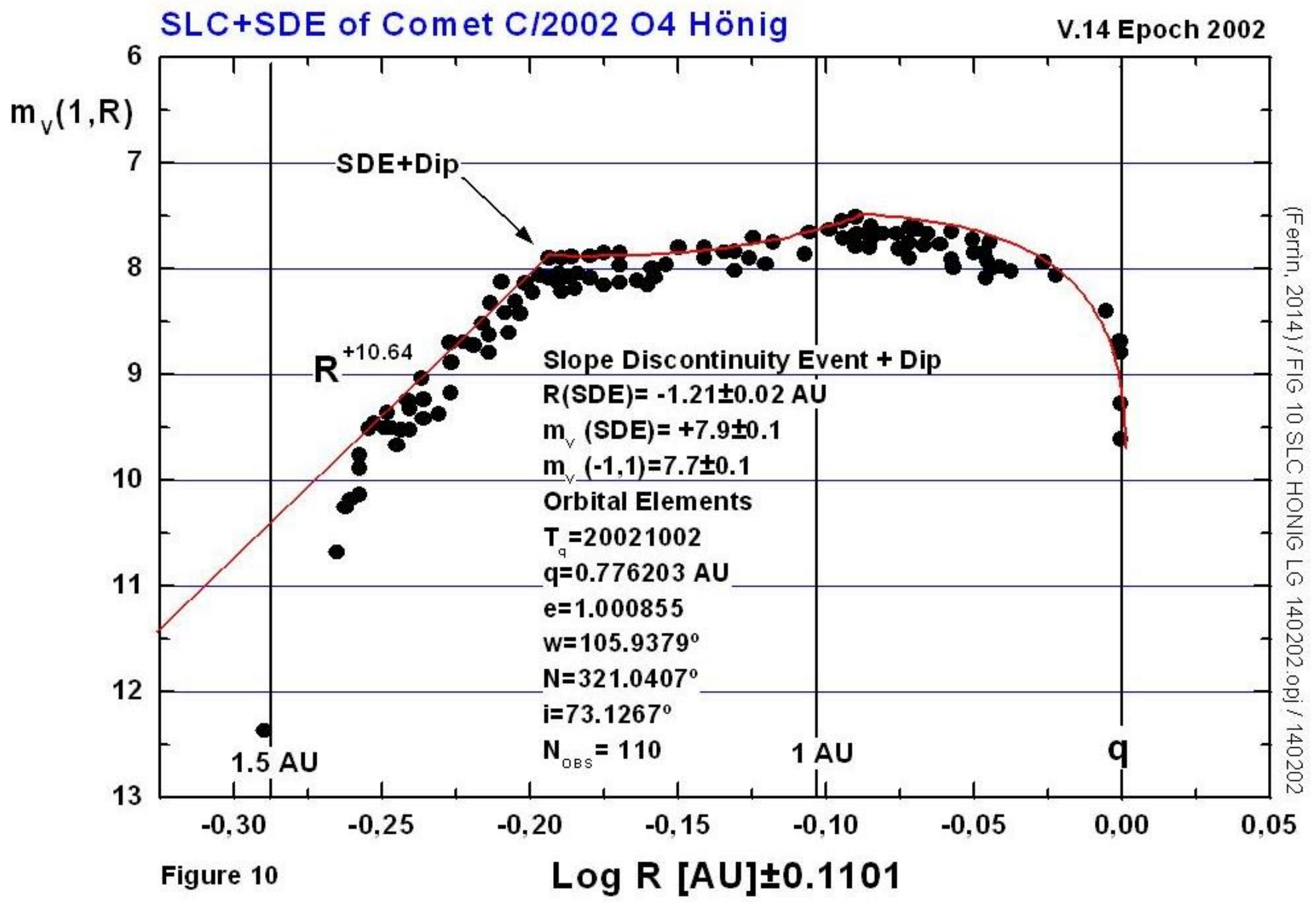

***Figure 10. The SLC of comet C/2002 O4 Hönig, log plot.** The comet exhibits a SDE+ near-standstill signature similar to the one exhibited by comet ISON. This comet disintegrated in a time span of 54 d after the SDE. The data for this plot comes from Sekanina (2002).*

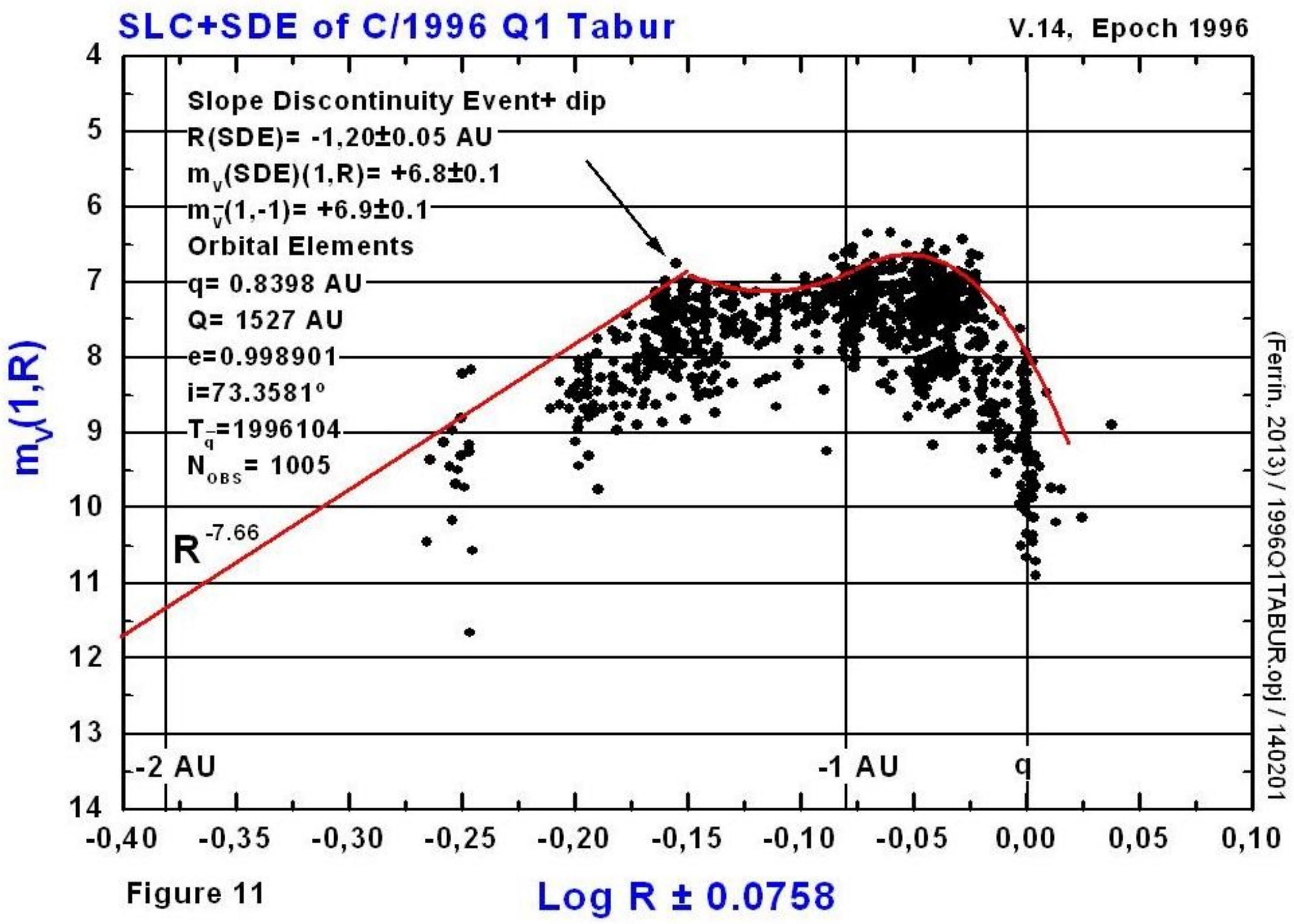

***Figure 11. The SLC of comet C/1996 Tabur*** *exhibits the same SDE+ near-standstill signature exhibited by comet Hönig suggesting that comet ISON might also disintegrate. The data for this comet comes from the ICQ dataset.*

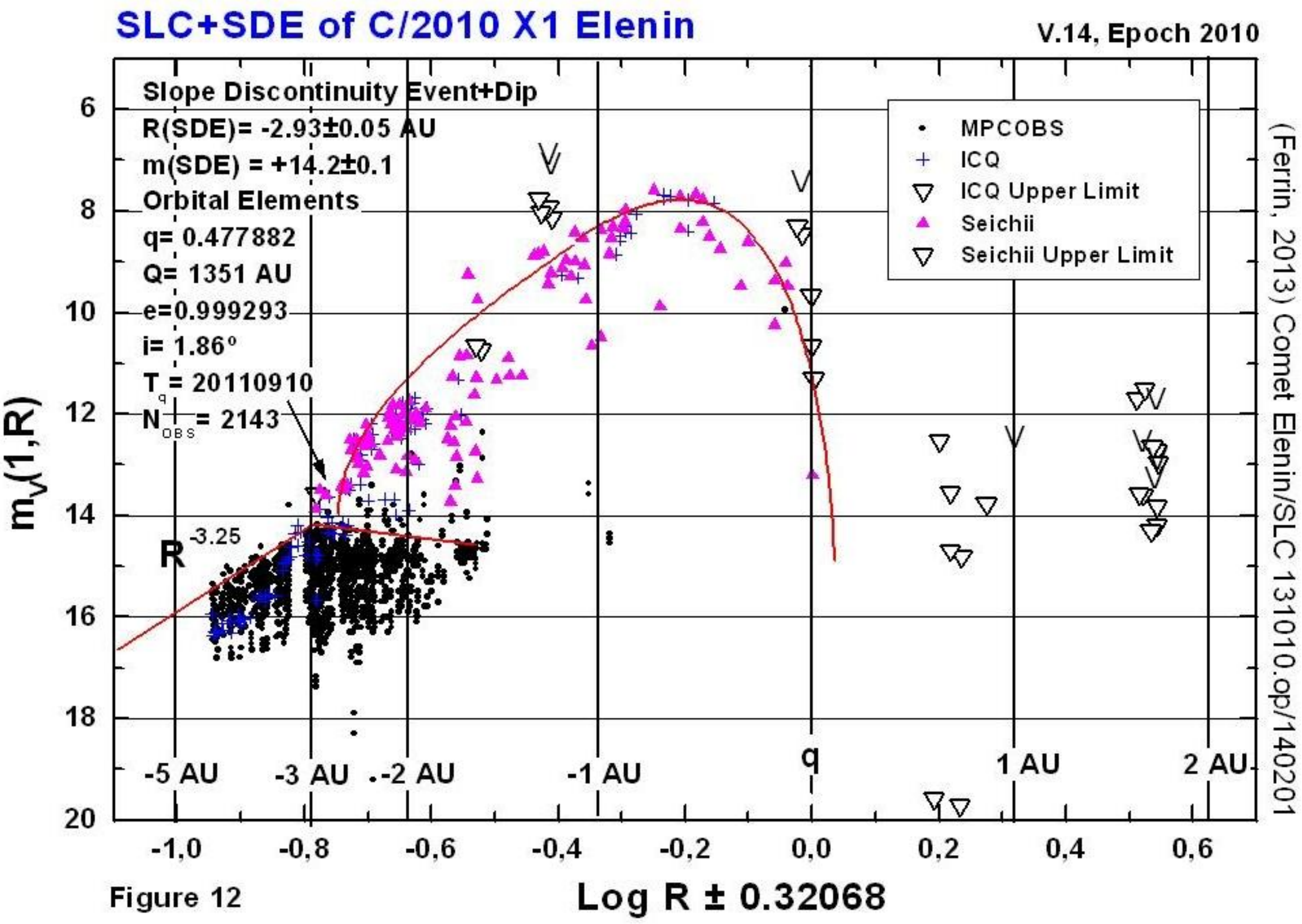

***Figure 12. The SLC of comet C/2010 X1 Elenin*** *exhibits the same SDE+ near-standstill signature exhibited by comets Hönig and Tabur and suggests that comet ISON could also disintegrate.*

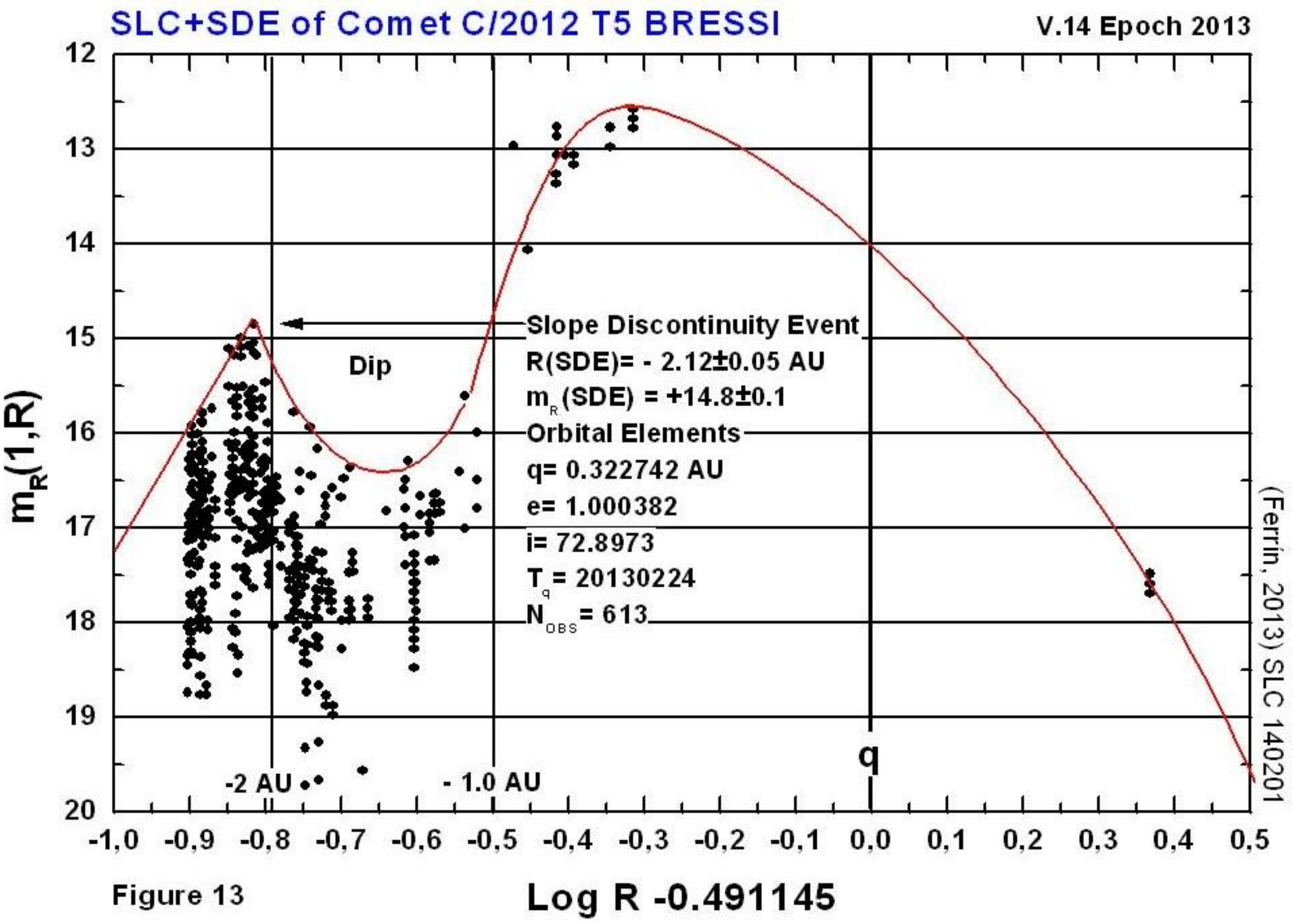

***Figure 13. The SLC of comet C/2012 T5 Bressi** exhibits the same SDE+dip signature exhibited by comets Hönig, Tabur and Elenin suggesting that comet ISON will also disintegrate. The data for this comet comes from the MPCOBS database.*

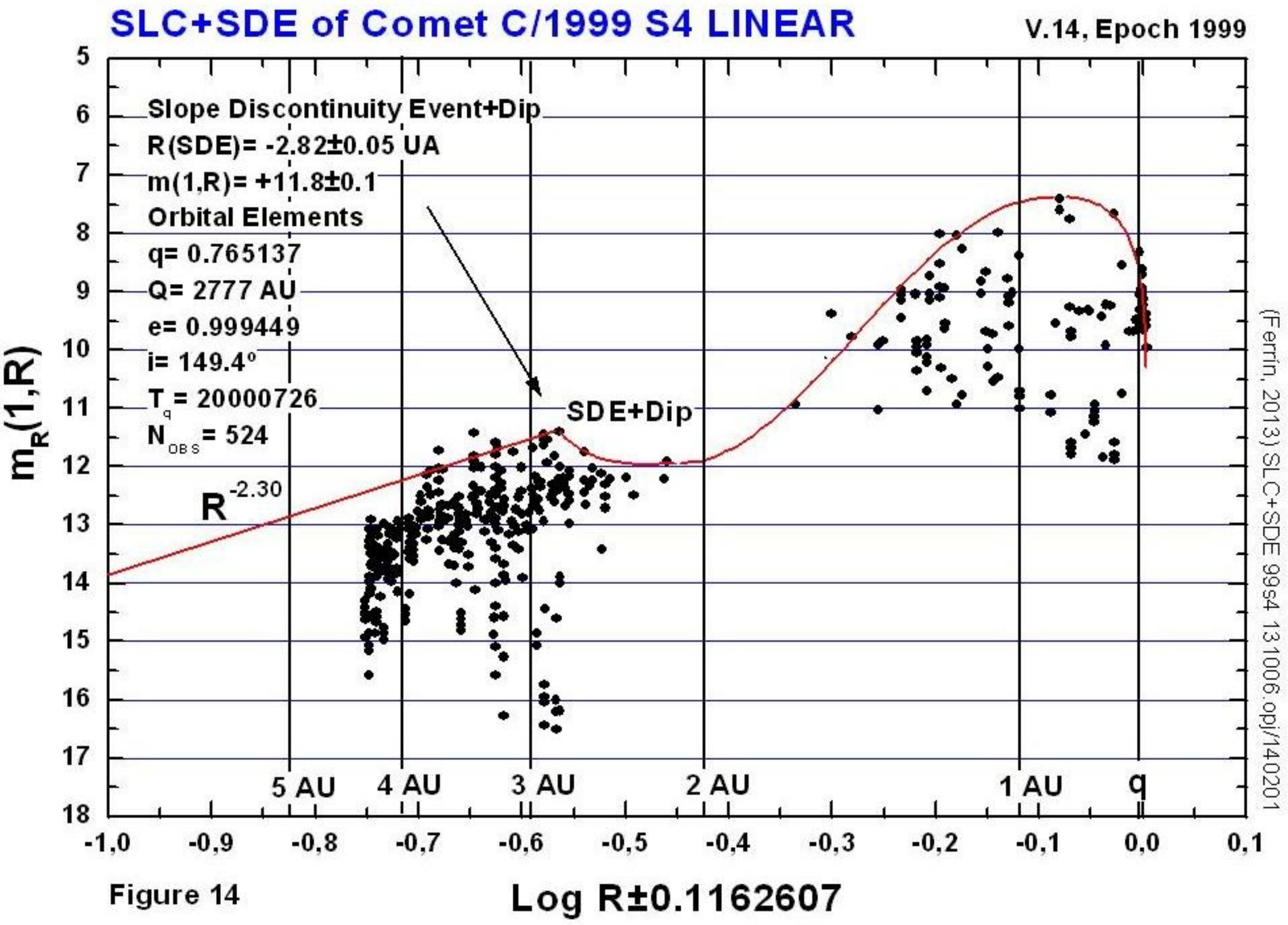

***Figure 14. Comet C/1999 S4 LINEAR*** *exhibits the same SDE+dip signature exhibited by comets Hönig, Tabur, Elenin and Bressi suggesting that comet ISON will most probably disintegrate. Data from MPCOBS database.*

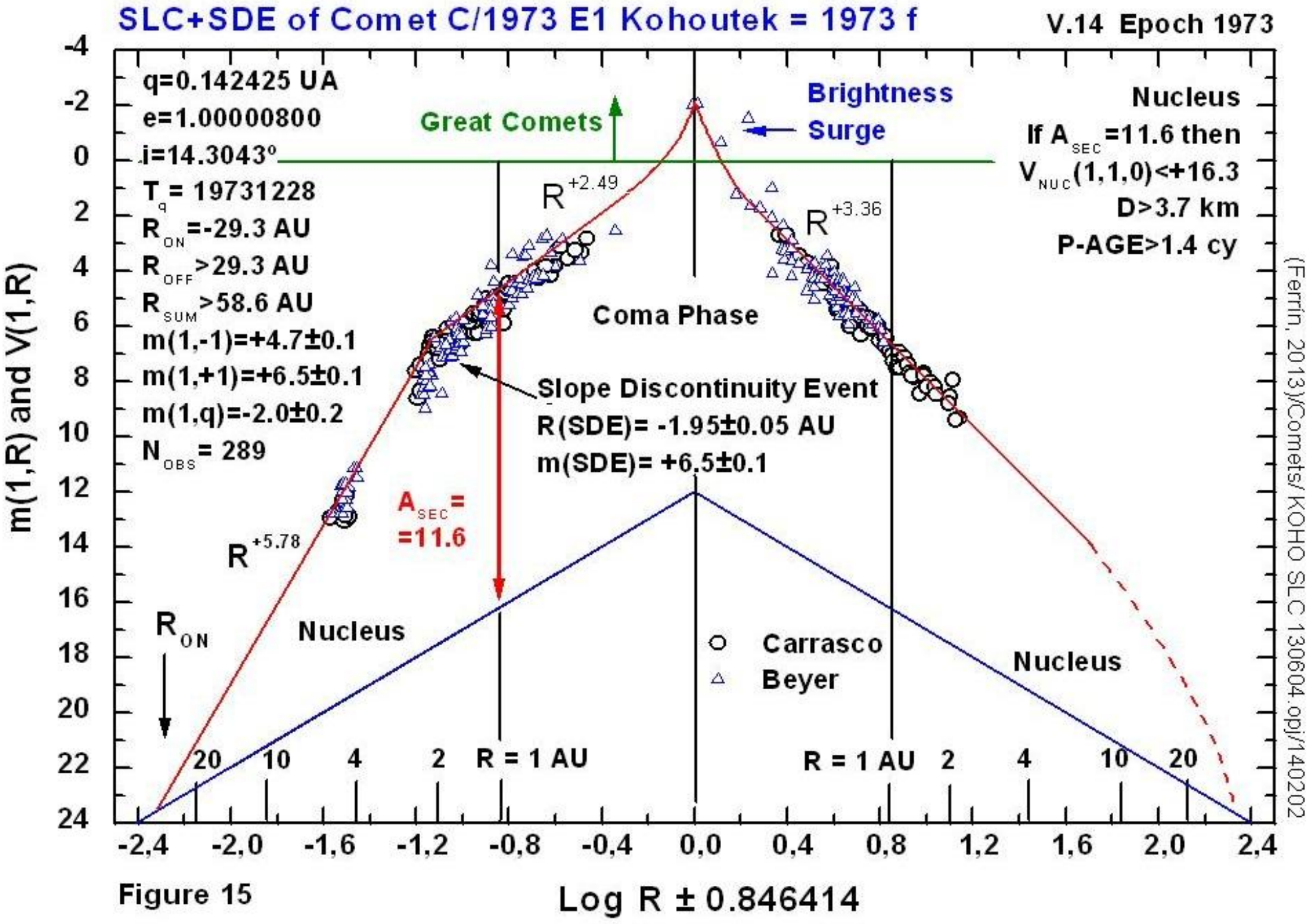

***Figure 15. The SLC of comet C/1973 E1 Kohoutek, the famous comet that was erroneously said to have fizzled.*** *This is an entirely normal SLC typical of Oort Cloud comets (confirmation the* Atlas I*). The comet was discovered with a power law* $R^{+5.78}$ *which would have produced a very bright comet at perihelion, were not for the SDE at R(SDE)= -1.95 ±0.05 AU,* $m_V$*(SDE)= +6.5±0.1. Even so the comet reached magnitude* $m_V(1,q) = -2.0$ *at perihelion, giving it entrance to the Great Comet Category (comets with negative magnitude at perihelion). So it did not really fizzle. The nucleus line in the form of a pyramid has been drawn assuming that* $A_{SEC} = 11.6$*, a maximum limit found for other comets (see Figure 1). The data for this plot comes from Carrasco (1990) and Beyer (1972).*

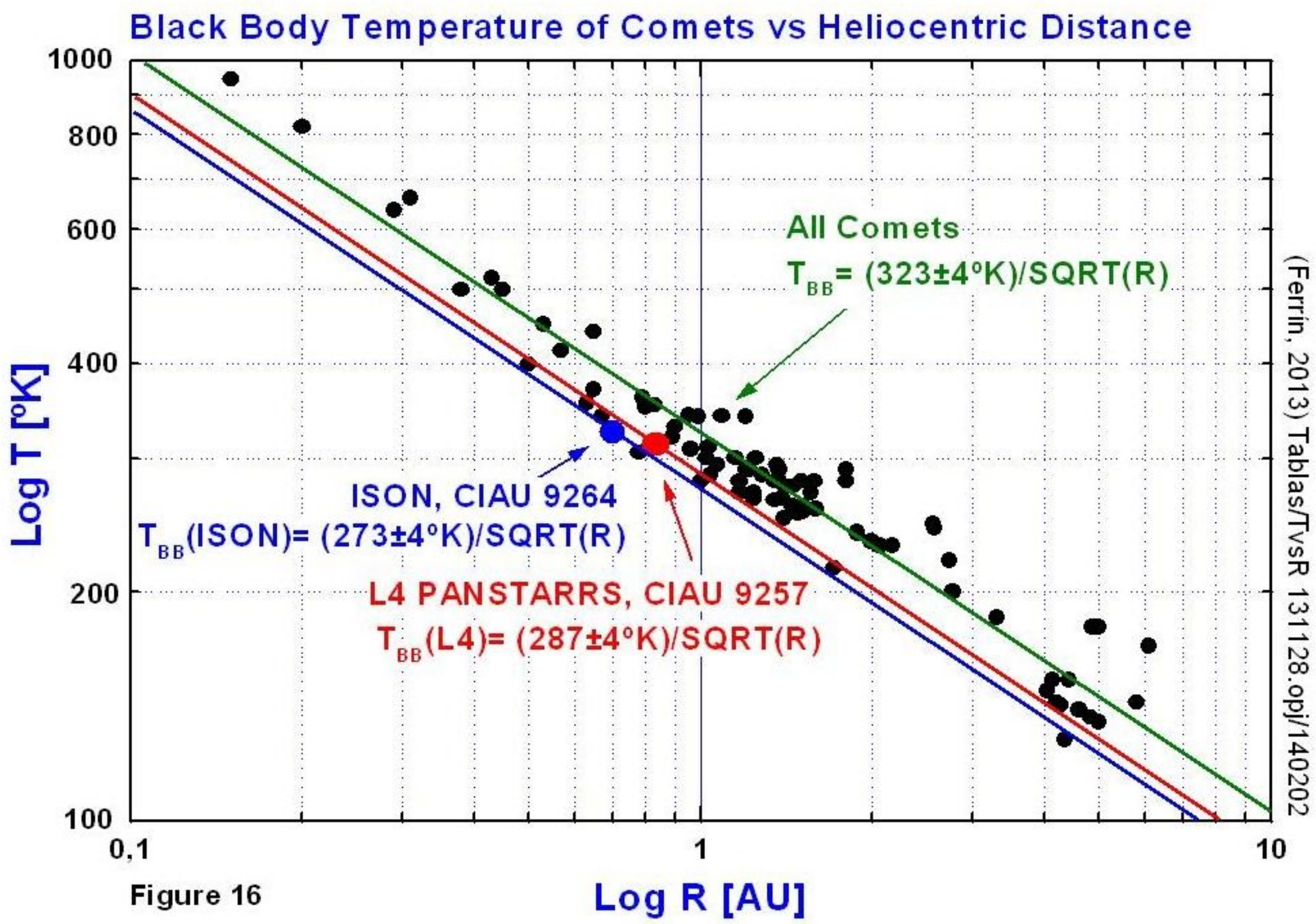

***Figure 16. Black Body (color) temperatures of comets** C/2011 L4 Panstarrs and C/2012 S1 ISON. The data comes from IAUC 9257 and 9264. The data for other comets has been compiled by Ferrín (2006).*

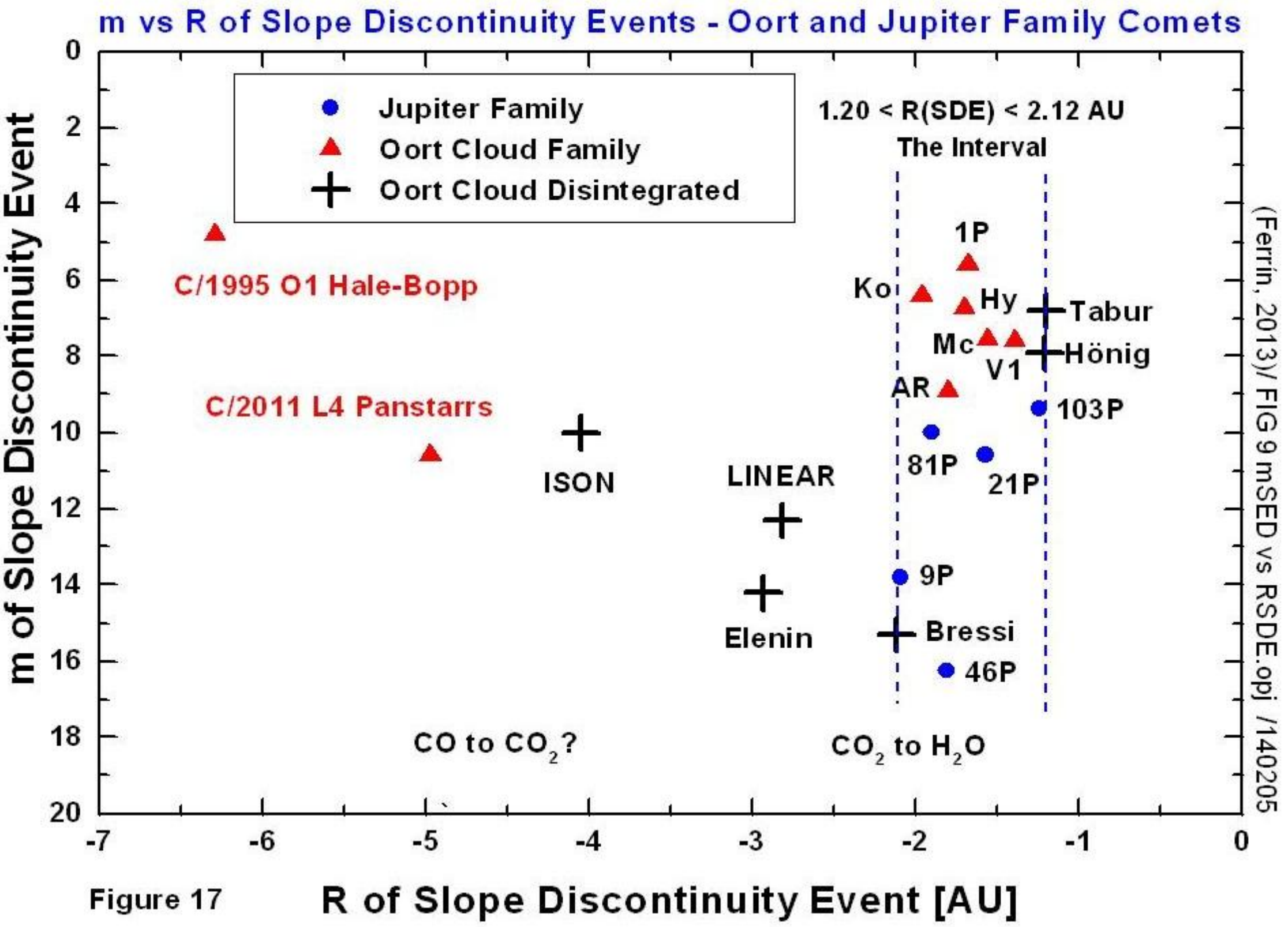

***Figura 17. The Slope Discontinuity Magnitudes of 19 comets*** *are plotted versus their slope discontinuity event distances from Table 2. 14 of 19 (74%) lie in a narrow vertical interval centered at R(Interval)= -1.68±0.07 AU and 1.20 < R(SDE) < 2.12 AU. Inside this* Interval *there are 5 Oort Cloud comets, 5 JF comets and 3 disintegrating comets. The interval has a width of 0.85 AU, while typical measuring errors are ±0.07 AU. Thus individual differences appear significant. The* Interval *may represent the changeover from CO2 to H2O controlling of the surface sublimation. It is not clear if the other five comets represent the CO to CO2 change over or the amorphous to crystalline ice transition (Prialnik and Bar-Nun, 1987) or both.*

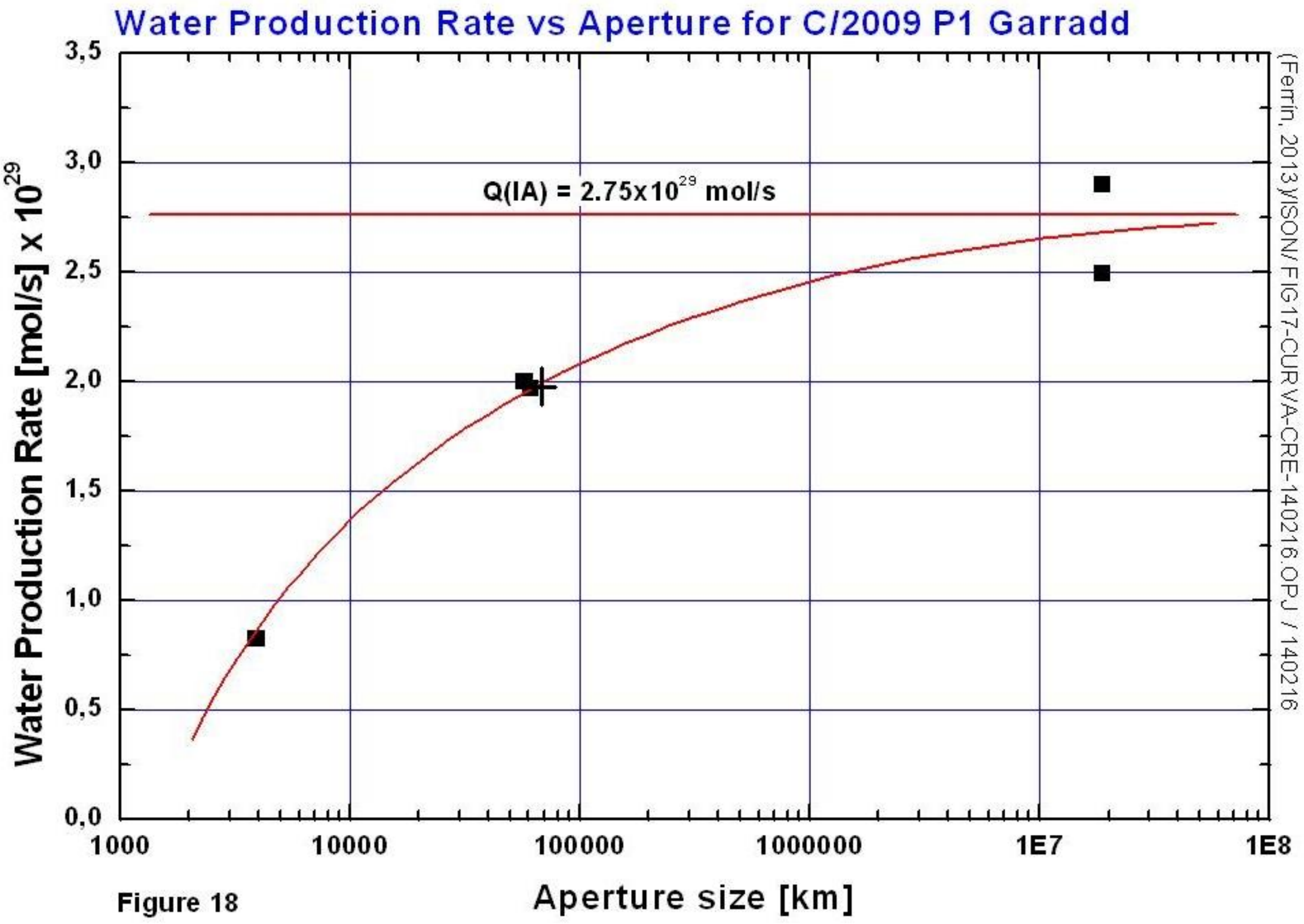

***Figure 18. The water production rate in comet C/2009 P1 Garradd*** *is shown near 2 AU pre-perihelion, as a function of aperture size (data from Combi et al., 2013 a). The flux increases asymptotically as aperture increases, allowing the definition of infinite aperture water production rates. This is one reason why it is advisable to adopt the envelope of the water production rate measurements, as the correct interpretation of the data.*

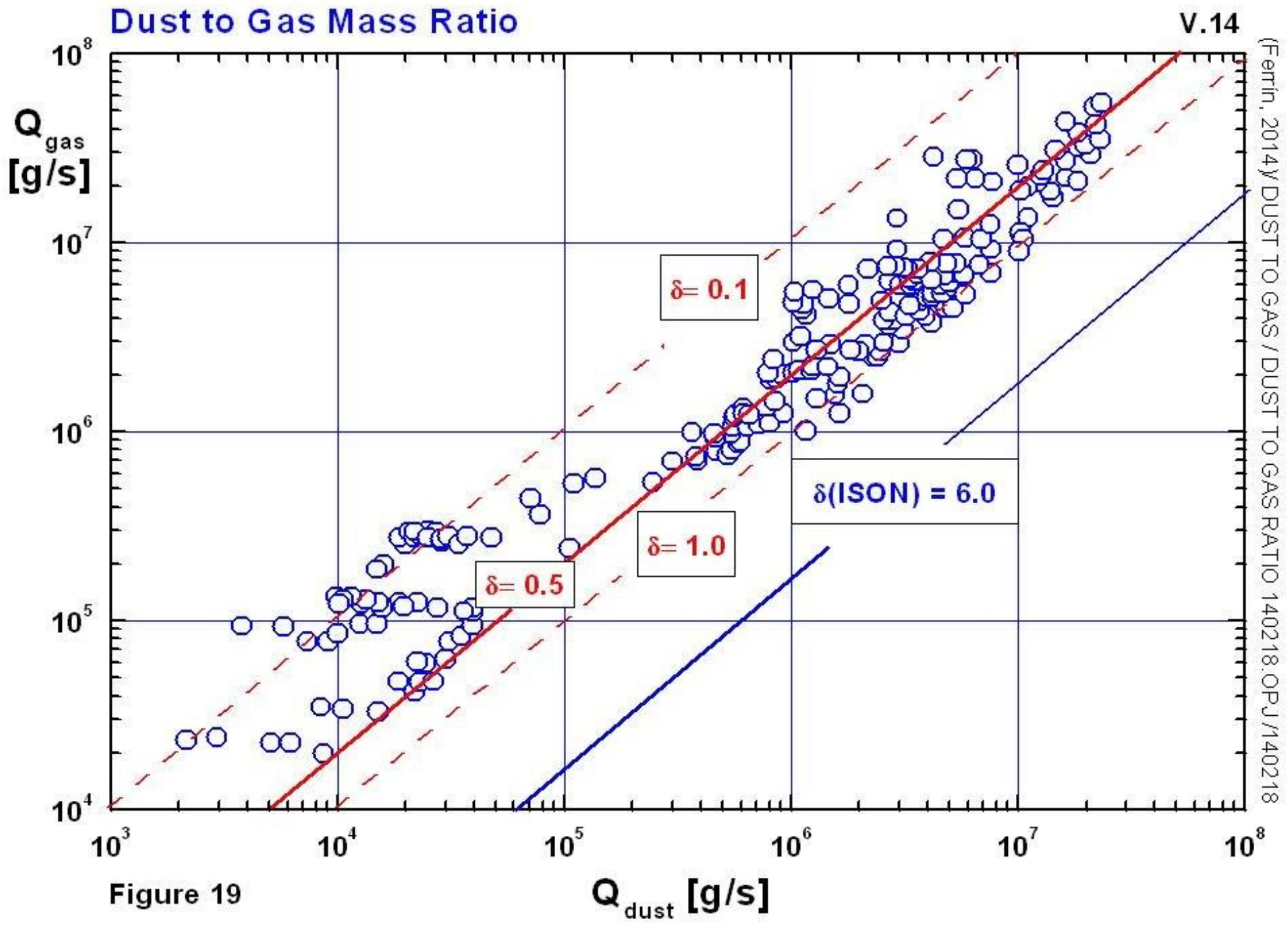

***Figure 19. The Dust to Gas Mass Ratio, δ.*** *The data for 4 comets (1P, 46P, 67P, and C/1996 B2), from de Almeida et al. (2009), shows that the dust to gas mass ratio, δ, is constrained to 0.1 < δ < 1.0. The ratio δ = 0.5, describes quite well the general tendency over 5 orders of magnitude, and it is adopted in this work to calculate the mass lost of several comets in Table 4. However, Figure 26 will show that the location of a comet in the RR versus ML-AGE diagram, is insensitive to the δ ratio. The mean dust to gas ratio for the apparition of comet ISON is shown and equals 6.0 implying that this was a very dusty comet.*

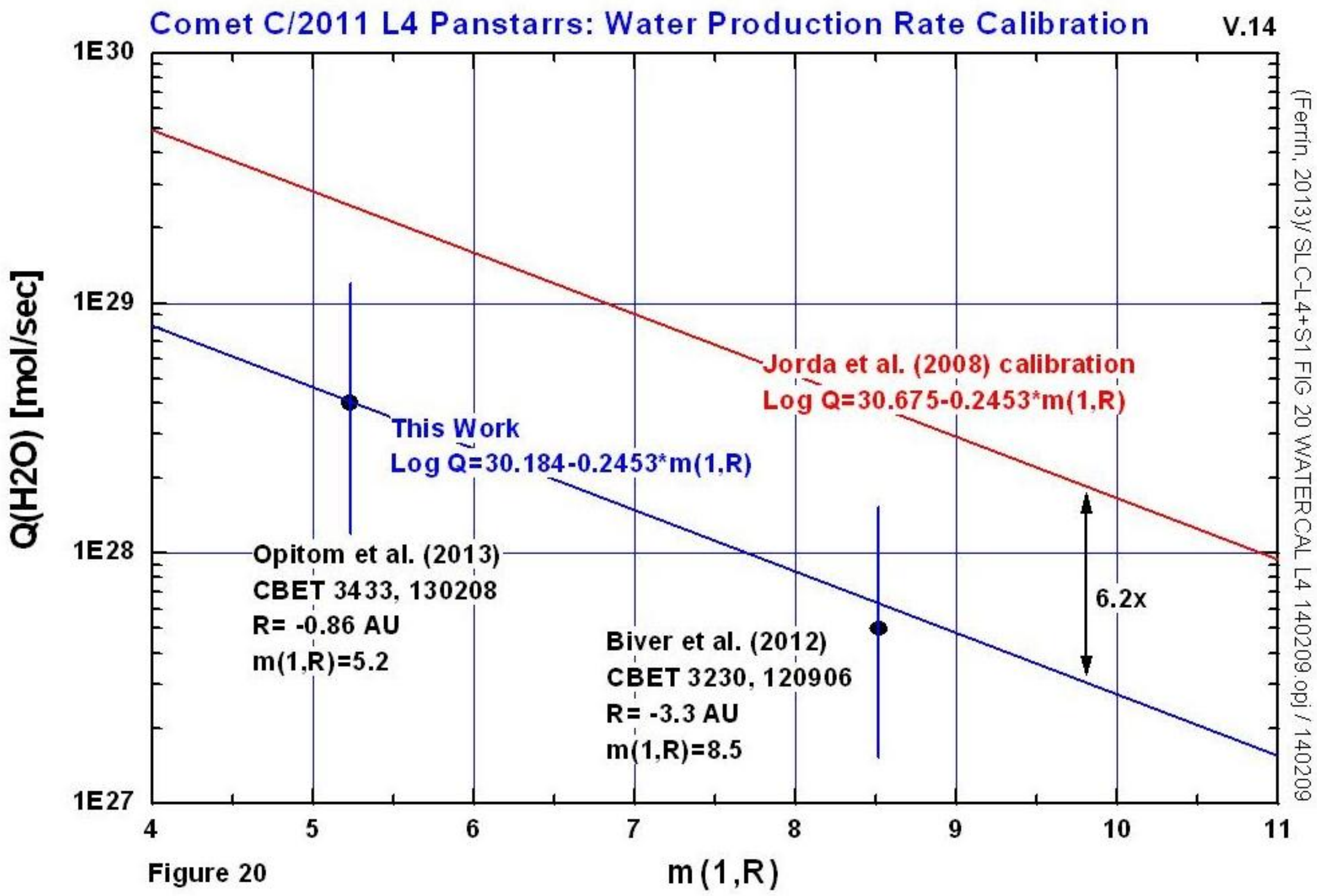

***Figure 20. Water calibration of comet C/2011 L4 Panstarrs.*** *Only two data points are available, however they agree quite well. The water production rate by Jorda, Crovisier & Green (2008) has been scaled down, but the slope has been preserved. This information will be used to calculate the water budget of this comet.*

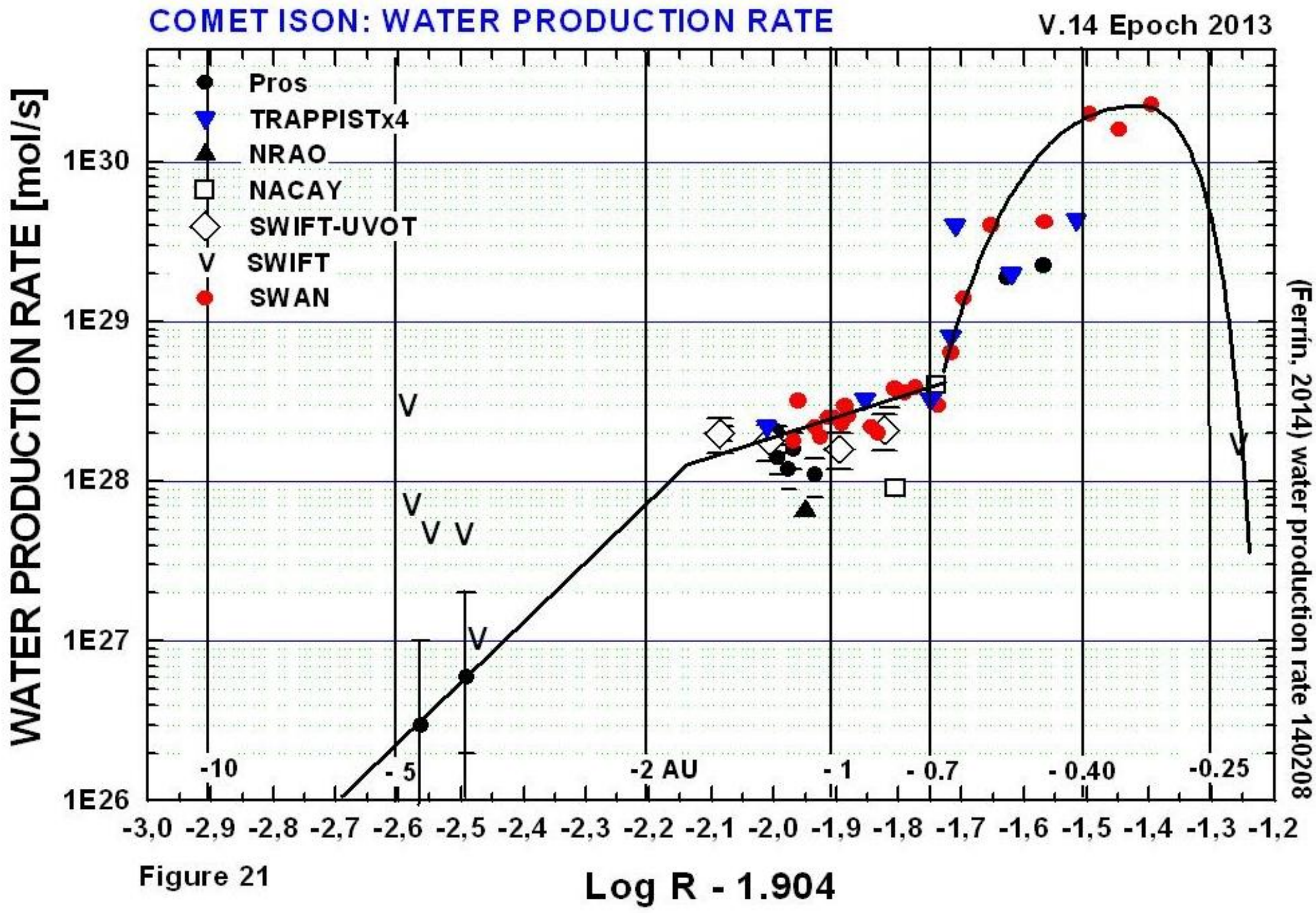

***Figure 21. Water Production Rate of comet C/2012 S1 ISON.*** *This plot can be converted to a time plot and the whole production rate can be integrated to find the water budget of the comet. We find a Water Budget WB= $3.94x10^{10}$ kg. The first two water data points are due to Schleicher (2013), while the data points of the outburst are due to Combi et al. (2013 a). Other data points come from Weaver et al. (2013), Dello Russo et al. (2013), Opitom et al. (2013), Crovisier et al. (2013), Bodewits et al. (2013 a, b), Bonev et at. (2013), Keane et al. (2013), Mumma et al. (2013).*

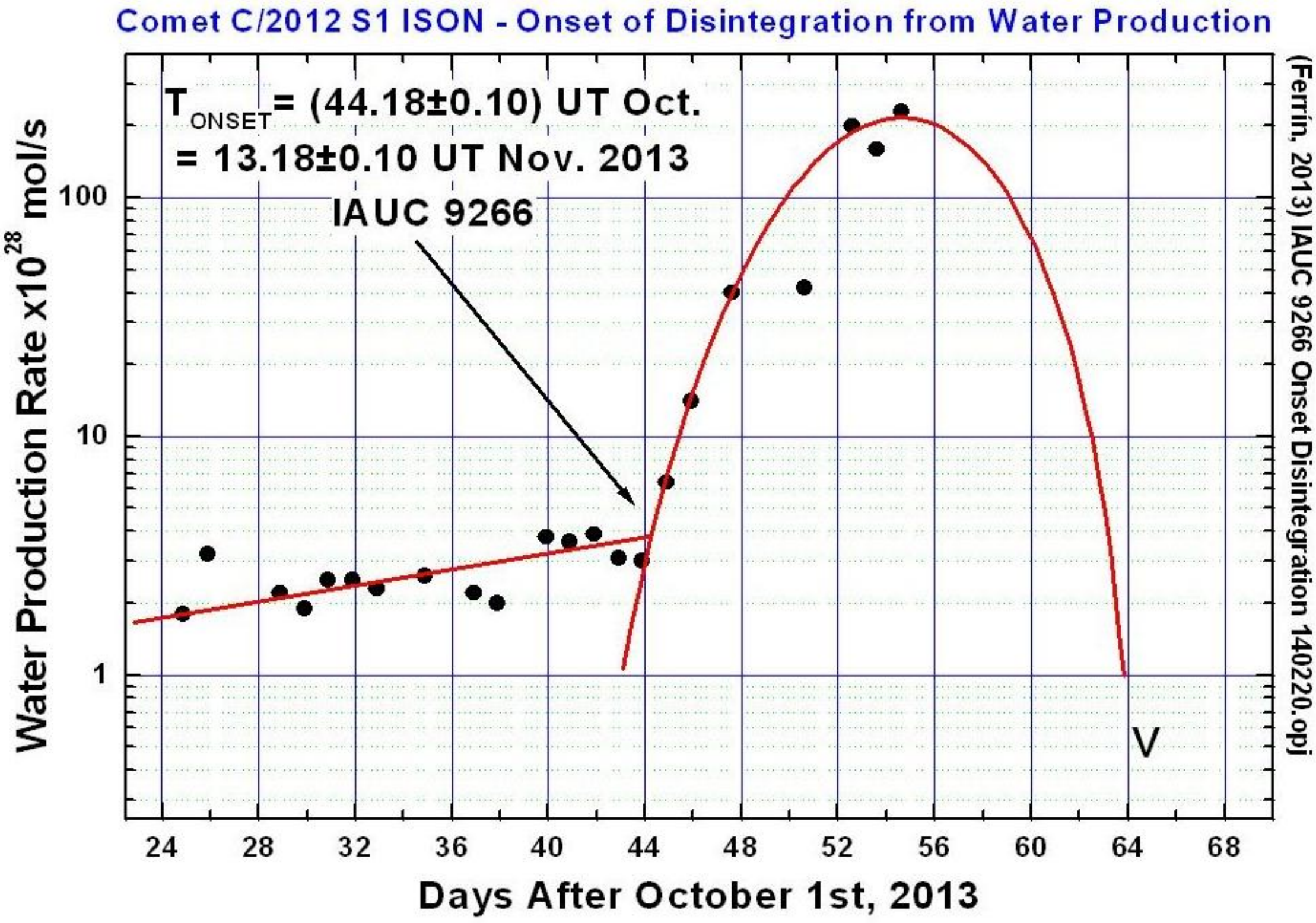

***Figure 22. The water production rate measured by Combi et al. (2013 a)*** *allows a precise determination of the onset of disintegration. We obtain* $T_{ONSET}$ *= 13.18± 0.10 UT Nov. 2013, at R= -0.68 ±0.01 AU. Compare this value with the one obtained independently in Figure 9 from visual photometry. The agreement is excellent.*

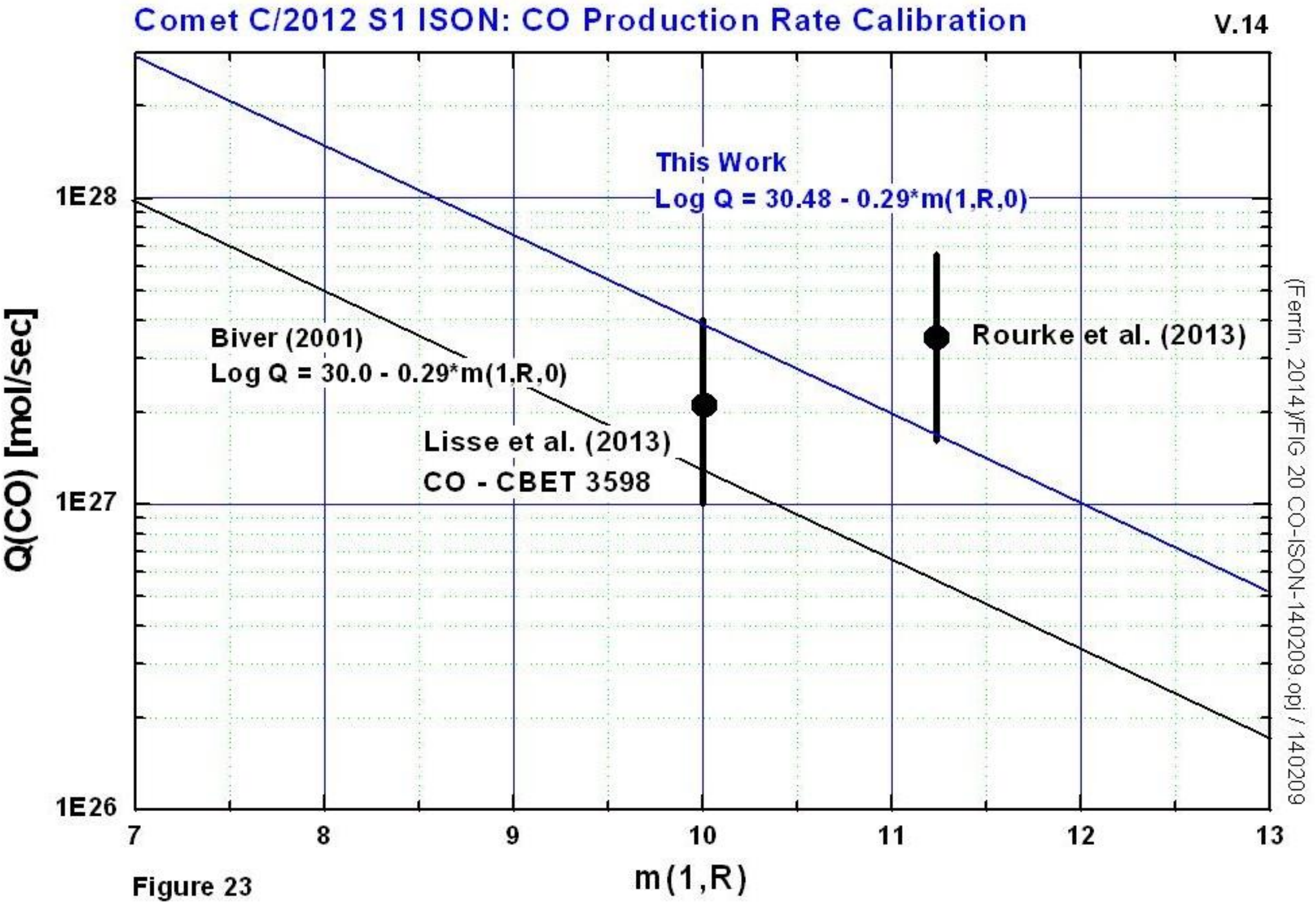

***Figure 23. The CO production rate of comet ISON.*** *Since there are only two data points, we had to use the calibration from Biver (2013). Plotting this calibration on a time plot and integrating up to the SDE (Li et al. 2013), we find a CO-Budget of 2.5x10*$^{9}$ *kg.*

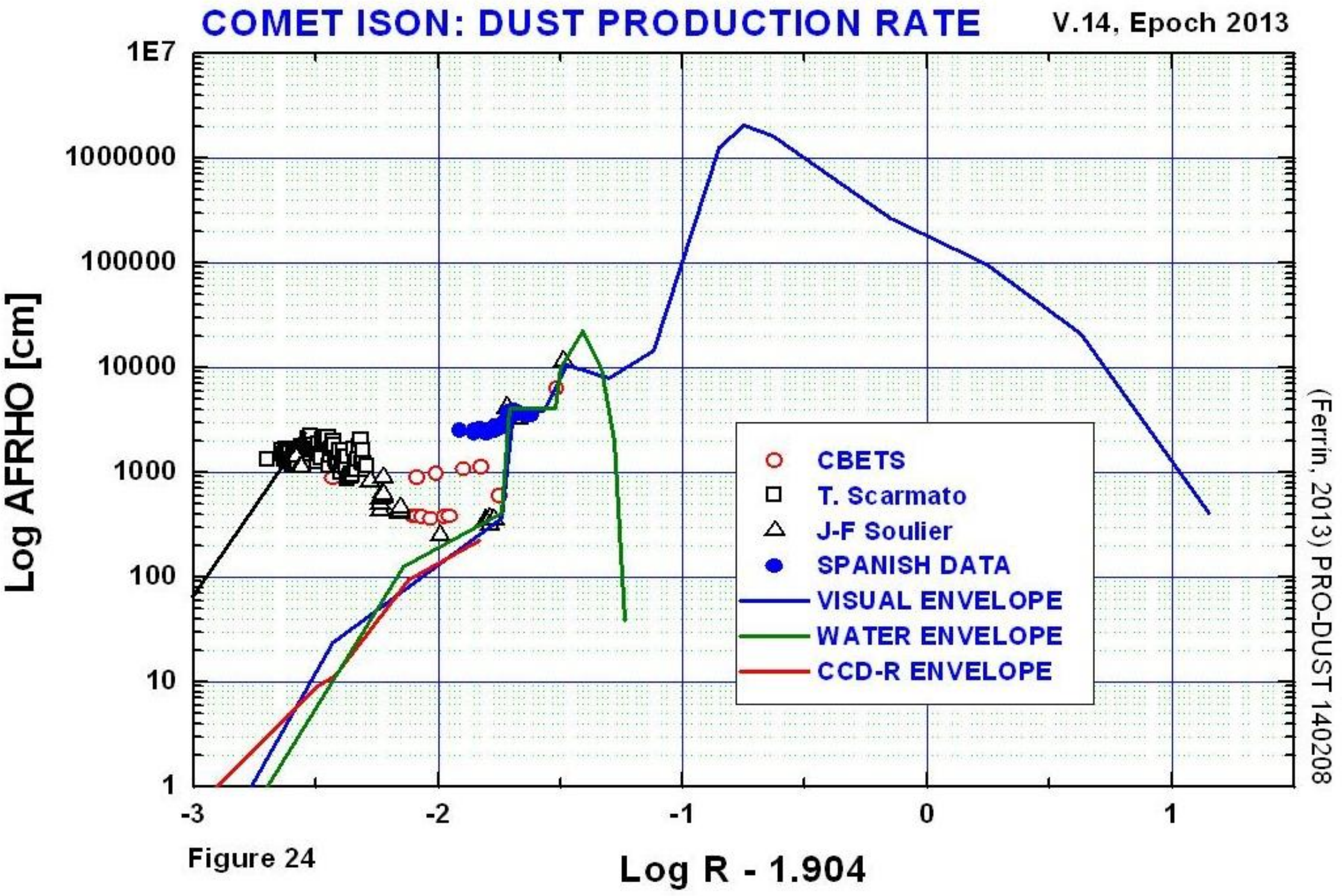

***Figure 24. Dust production rate for comet C/2012 S1 ISON.*** *Data comes from Lisse et al. (2013), Opitom et al. (2013), Bodewits et al. (2013 a, b), Scarmato (2013). Since there is a lack of data near perihelion, we have scaled the visual data to be able to integrate the whole apparition. Near perihelion the gas was exhausted and the brightness was only due to the dust.*

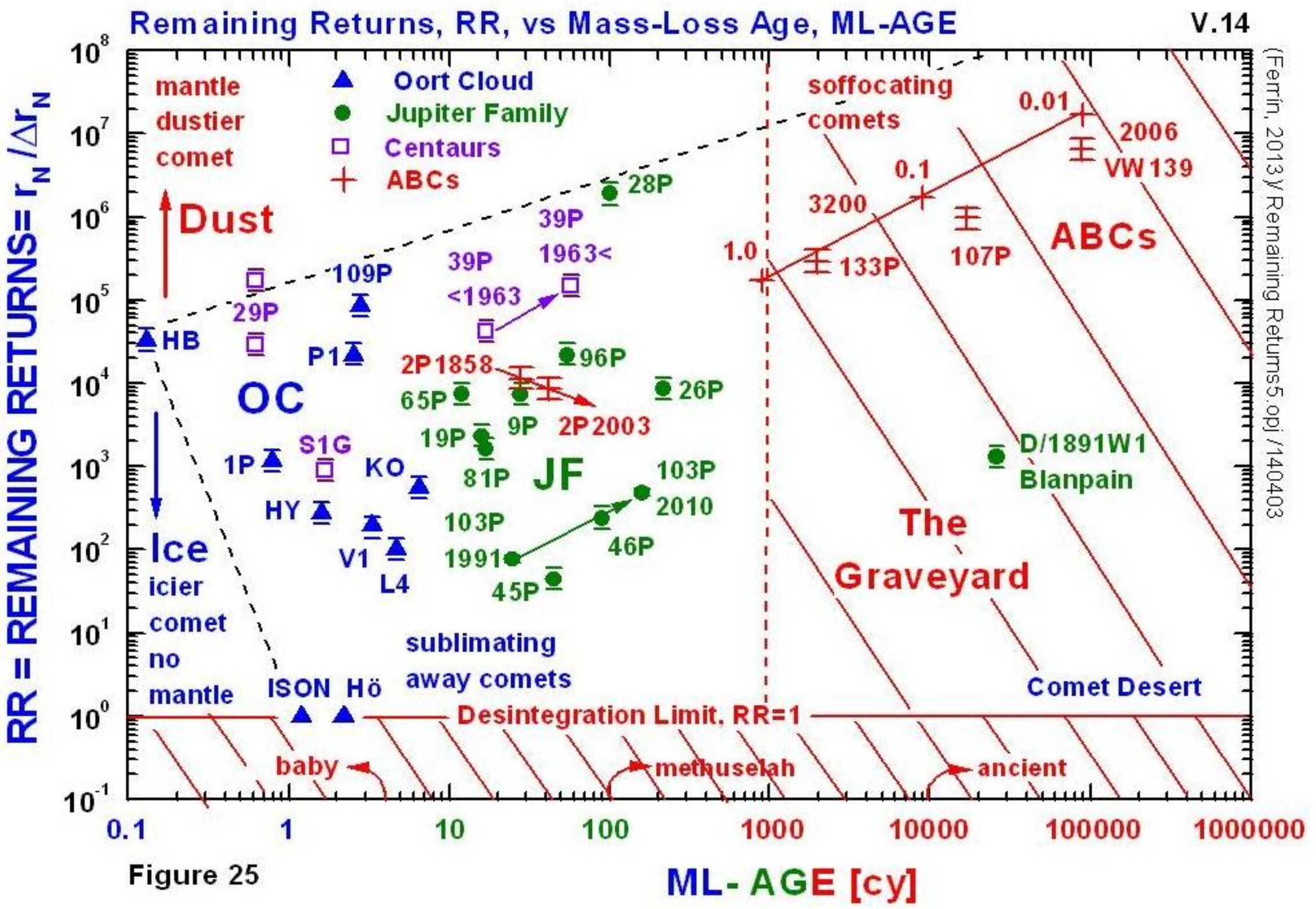

***Figure 25. Remaining Returns versus Mass-Loss Age diagram.***

*Nomenclature: 1P=1P/Halley. KO=C/1973 E1 Kohoutek. P1=C/2009 P1 Garradd. V1=C/2002 V1 NEAT. HB=C/1995 O1 Hale=Bopp. L4=C/2011 L4 Panstarrs. HY=C/1996 B1 Hyakutake. Hö=C/2002 O4 Hönig. S1G = P/2009 S1 Gibbs. The calculation is done for the ratio dust/gas δ = 0.5. ABC = Asteroidal Belt Comet. OC = Oort Cloud. JF = Jupiter Family.*

*(1) The diagram covers a large area: 8 orders of magnitudes in the vertical, 6 in the horizontal axis.*

*(2) If a comet were made of pure ice, the layer removed by apparition would remain approximately constant (see demonstration in Section 9), and r/Δr would tend to zero as the comet sublimates away. If the comet contained much dust, part of it would remain on the surface, Δr would tend to zero and r/Δr would tend to infinity. Thus sublimating away comets move down, suffocating comets move up in the diagram.*

*(3) The location of a comet is not sensitive to the dust to gas mass ratio, δ. The error bars represent the limits 0.1< δ<1.0.*

*(4) Four comets in the graveyard are being suffocated: 107P, 133P, 3200 and 2006 VW139.*

*(5) 2006 VW139 is the most extreme object in the upper right hand corner.*

*(6) Five comets belong to the graveyard in this definition (1000 cy < P-AGE), 107P, 133P, D/1891W1 Blanpain, 2006 VW139 and 3200 Phaeton. The location of 3200 is plotted for three diameters of the dust particles, 1.0 mm, 0.1 m and 0.01 mm. It is not surprising that ABCs occupy the upper right hand corner of the diagram. This is expected on physical grounds. The diagram says that they are old (large ML-AGE) and that they have a substantial dust or crust layer (large r/Δr ).*

*(7) The diagram separates comets into classes: Oort Cloud comets on the left, Jupiter Family comets in the middle, asteroidal belt comets, ABCs, above right, and disintegrating comets below on the RR=1 line.*
*(8) The 39P jump. Comet 39P had a close encounter with Jupiter in 1963 and the orbit changed substantially. Fortunately we know the SLC quite well (Paper II), and not only it is possible to estimate the old mass loss but also estimate the future mass loss. The comet jumps in the diagram due to the orbital change.*
*(9) Comet 2P/Encke shows evolution of the parameters between the 1858 and the 2003 apparitions. Comet 103P shows evolution between the 1991 and the 2010-11 apparitions (Ferrín et al., 2012). Comet 39P jumped in position due to an orbital change. Comets evolve from the left side toward the right side as they age. If they move up they are choked by a dust crust. If they move down they sublimate away. If they move toward the left, they rejuvenate. If they are in the graveyard and move toward the left, they become Lazarus Comets (Ferrín et al. 2013a). If they approach RR=1, they disintegrate. If they are scattered, they jump. Thus RR versus ML-AGE, is an evolutionary diagram.*
*(10) There should be a* comet desert *in the lower right hand side of the diagram. We expect comets to sublimate away in a time scale much shorter than would take for a comet to suffocate, thus we do not expect to find comets sublimating away in that region. No objects are found plotted in that area as expected.*
*(11) Since the diagram is log-log and covers 8 and 6 orders of magnitude in the Y and X axis, the diagram is very forgiving. A factor of 2 error in any of the two variables changes the location of the data point by less than an order of magnitude.*
*(12) RR=1 is the disintegration limit. The comets on the disintegration limit are C/2002 O4 Hönig, which in fact disintegrated (Sekanina, 2002) and ISON.*
*(13) Comet ISON is the youngest of the Oort Cloud group.*
*(14) Another way to think about this diagram is that the Mass Loss Age is the inverse of the Water Budget, and the WB depends on the product of the surface area times the production rate per unit area. Thus large comets tend to be on the left, and small comets tend to be on the right. The plot segregates the comets by decreasing size from left to right. And segregates them by mode of evolution (suffocation above), sublimation (below). Since 90% of these comets have 1.0 < D < 10 km, then diameter accounts for a factor of ~100. Since the horizontal axis varies by a factor of $10^6$ it is concluded that evolution accounts for a factor of $10^4$ .*

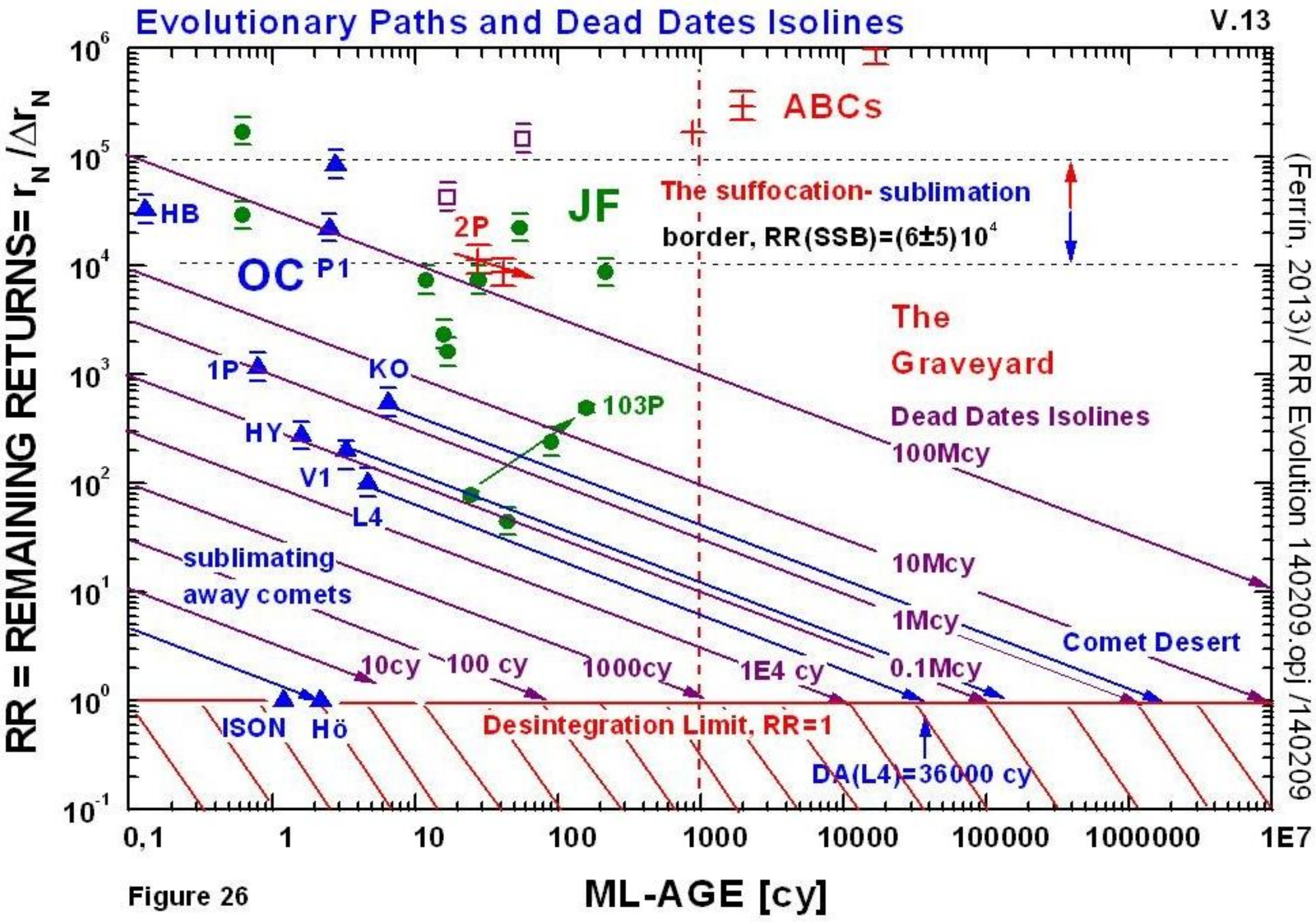

***Figure 26. Evolutionary Paths in the RR vs ML-AGE diagram.*** *In a simple model applied to sublimating comets, the layer removed from the cometary nucleus is constant as a function of time (see Section 9). Then the evolutionary lines are straight lines of negative slope. These lines will intercept the RR = 1 line, the disintegration line, at the Death Age, DA. The death ages of several comets have been measured. We find, DA(KO) = 1.7E06 cy, DA(V1) = 1.4E05, DA(L4) = 36000cy, DA(Hö) = 6 cy. Comet 2P/Encke is following an isoline, but comet 103P/Hartley 2, is not. The model predicts DAs in the desert region. No objects have been found in this region (0/27 or <3.6%). Since sublimating comets move down and suffocating comets move up, there must be a region where comets move horizontally. This is the location of the suffocation-sublimation-border (SSB). Its location is estimated at RR(SSB) = (6±5) $10^4$. Since the error is large, it is not known on what side of the border comets Hale-Bopp or C/2009 P1 Garradd are located.*

Table 1. Slope Discontinuity Event of comet C/2012 S1 ISON, measured from different data bases.

| DATABASE-------> PARAMETER | CCD-R | MPCOBS | MULTI-APERTURE |
|---|---|---|---|
| T(SDE) | 2013-04-14 | 2013-04-09 | 2013-04-21 |
| Δt (SDE)= t - $T_q$ | -228±2 | -233±1 | -221±1 |
| R(SDE)(AU) | -4.10±0.03 | -4.16±0.03 | -4.02±0.02 |
| $m_R(\Delta,R)$(SDE) | 11.4±0.1 | ------------ | 11.8±0.1 |
| $m_V(1,R)$(SDE) | ------------ | 11.4±0.1 | ----------- |

Table 2. Distances, magnitudes at Slope Discontinuity(SDE) q,e,i,Tiss.

| R (SDE) | $m_V$ (SDE) | q | 1/a-orig or e | i | Tiss | Comet |
|---|---|---|---|---|---|---|
| **-1.21** | **7.9** | **0.776** | **-0.000772** | **73.1** | **0.31** | **C/2002 O4 Hönig** |
| **-2.12** | **15.3** | **0.322** | **-0.000227** | **72.9** | **0.201** | **C/2012 T5 Bressi** |
| **-2.82** | **12.3** | **0.765** | **-0.000055** | **149.4** | **-0.93** | **C/1999 S4 LINEAR** |
| -1.8 | 8.9 | 0.316 | -0.000020 | 119.9 | 0.00 | C/1956 R1 Arend-Roland |
| **-4.05** | **10.0** | **0.012** | **0.000009** | **62.1** | **0.00** | **C/2012 S1 ISON** |
| -1.96 | 6.4 | 0.142 | 0.000020 | 14.3 | 0.00 | C/1973 E1 Kohoutek |
| -7.18 | 13.6 | 1.398 | 0.000029 | 129.0 | ---- | C/2013 A1 Siding Spring |
| -4.97 | 10.6 | 0.302 | 0.000030 | 84.3 | 0.00 | C/2011 L4 Panstarrs |
| -1.56 | 7.6 | 0.171 | 0.000031 | 77.1 | 0.00 | C/2006 P1 McNaught |
| **-3.22** | **12.7** | **0.478** | **0.000111** | **1.9** | **0.86** | **C/2010 X1 Elenin** |
| -1.7 | 6.7 | 0.230 | 0.001546 | 124.9 | -0.33 | C/1996 B2 Hyakutake |
| **-1.20** | **6.9** | **0.840** | **0.001826** | **73.0** | **0.34** | **C/1996 Q1 Tabur** |
| -1.39 | 7.6 | 0.099 | 0.002297 | 81.7 | 0.06 | C/2002 V1 NEAT |
| -6.29 | 4.8 | 0.914 | 0.003805 | 89.4 | 0.04 | C/1995 O1 Hale-Bopp |
| -3.47 | 8.5 | 1.550 | 1.001007 | ---- | ---- | C/2009 P1 Garradd |
| -1.68 | 5.6 | 0.586 | 0.967943 | 162.3 | -0.61 | 1P/Halley |
| -1.57 | 10.6 | 1.031 | 0.707045 | 31.9 | 2.46 | 21P/Giacobinni-Zinner |
| -1.24 | 9.4 | 1.059 | 0.694533 | 13.6 | 2.64 | 103P/Hartley 2 |
| -1.8 | 16.3 | 1.057 | 0.659295 | 11.7 | 2.81 | 46P/Wirtanen |
| -1.9 | 10.0 | 1.598 | 0.537385 | 3.2 | 2.88 | 81P/Wild 2 |
| -2.09 | 13.8 | 1.509 | 0.516946 | 20.5 | 2.90 | 9P/Tempel 1 |

* JF-Box,-1.20<R(SDE)<-2.12 AU or <R>= -1.70±0.08 AU, N=12
Disintegrated comets are highlighted in black.

Table 3. Absolute magnitudes, P-AGEs and power laws

| Comet | P-AGE[3] (cy) | $m_{V-TOT}$ (1,-1) | n in R^n PreSDE | n in R^n PostSDE |
|---|---|---|---|---|
| Oort Cloud | | | | |
| ***C/2002 O4 Hönig*** | ***D[1]*** | ***+7.9±0.5*** | ***-10.6*** | ***-1.40*** |
| C/1956 R1 Arend Roland | --- | +6.3±0.1 | -7.2 | -4.0 |
| C/2011 L4 Panstarrs | --- | +6.7±0.1 | -8.7 | -2.24 |
| C/2006 P1 McNaught | 11< | +5.2±0.1 | -10.6 | -1.55 |
| P/1973 E1 Koho | --- | +5.6±0.1 | -5.8 | -2.49 |
| ***C/2012 S1 ISON*** | ***D[1]*** | ***+12.2±0.1*** | ***-5.0*** | ***-0.34*** |
| C/2002 V1 NEAT | 15< | +6.7±0.1 | -13.0 | -3.37 |
| C/1996 B2 Hyak | 18 | +4.8±0.1 | -11.6 | -2.33 |
| ***C/1996 Q1 Tabur*** | ***D[1]*** | ***+6.9±0.1*** | ***-11.2*** | ***-1.28*** |
| C/1995 O1 HB | 2.4 | +0.6±0.1 | -10.7 | -2.58 |
| 1P/Halley | 7.1 | +3.9±0.1 | -8.9 | -3.35 |
| Jupiter Family | | | | |
| 21P | 22 | +8.0±0.1 | -9.1 | -5.16 |
| 103P | 14 | +8.3±0.1 | -9.5 | -5.55 |
| 46P | 15 | +7.6±0.1 | -5.2 | -7.20 |
| 81P | 13 | +5.8±0.2 | -9.3 | -7.03 |
| 9P | 22 | +6.4±0.2 | -7.7 | -6.50 |

(1) Disintegrating comets are in italics (D).

Table 4. Water budget, WB, water budget age, WBAGE, Remaining Returns, $RR=r_N/\Delta r_N$, δ=M(dust)/M(gas).WB-AGE[cy]=3.58E+11/WB [kg], WB/WB(1P)=WB/4.51E11 kg

| Comet | WB [kg] | WB-AGE [cy] | WB<br>----- %<br>WB(HB) | $r_N$ [km] | $\Delta r_N$[m] δ=0.5 | RR δ=0.1 | RR δ=1 | RR δ=0.5 |
|---|---|---|---|---|---|---|---|---|
| C/1995O1 Hale-Bopp | 2.67E+12 | 0.13 | 100 | 27 | 0.82 | 44600 | 24550 | 32732 |
| 29P/SW 1 | 5.60E+11 | 0.63 | 20.9 | 15.4 | 0.53 | 3.9E4 | 2.2E4 | 2.9E4 |
| 29P/SW 1 | 5.60E+11 | 0.63 | 20.9 | 27.7 | 0.16 | 2.3E5 | 1.3E5 | 1.7E5 |
| 1P/Halley | 4.51E+11 | 0.79 | 8.4 | 4.9 | 4.2 | 1580 | 868 | 1158 |
| C/1996B2 Hyakutake | 2.25E+11 | 1.6 | 8.4 | 2.4 | 8.8 | 370 | 204 | 272 |
| P/2011 S1 Gibbs | 2.14E+11 | 1.7 | 8.0 | 3.5 | 3.9 | 1213 | 667 | 889 |
| C/2002 O4 Hönig[4] | 1.50E+11 | 2.2 | 8.0 | 0.35 | 0.35 | ---- | 1 | ---- |
| C/2009 P1 Garradd | 2.14E+11 | 1.7 | 5.3 | 3.5 | 3.48 | ---- | 1.0E3 | ---- |
| 109P/Swift-Tuttle | 1.29E+11 | 2.8 | 4.8 | 13.5 | 0.15 | 115500 | 63500 | 84685 |
| C/2002 V1 NEAT | 1.10E+11 | 3.3 | 4.1 | 1.7 | 8.6 | 247 | 135 | 198 |
| C/2012S1ISON[3] | 5.2E+10 | 6.9 | 0.24? | 0.58 | ---- | ---- | ---- | 1 |
| C/2012S1ISON | 1.53E+11 | 2.4 | 0.16? | 0.41 | ---- | ---- | ---- | 1 |
| C/2011L4 Panstarrs | 7.60E+10 | 4.7 | 2.8 | 1.2 | 11.9 | 138 | 75 | 101 |
| C/1973E1 Kohoutek | 5.50E+10 | 6.5 | 2.1 | 1.9 | 3.4 | 755 | 415 | 553 |
| 65P/Churyumov-Gera | 3.06E+10 | 12 | 1.1 | 3.7 | 0.50 | 10020 | 5500 | 7349 |
| 19P/Borrelly | 2.17E+10 | 16 | 0.81 | 2.25 | 0.96 | 3178 | 1748 | 2330 |
| 39P/Oterma <1963 | 2.12E+10 | 17 | 0.80 | 3.20 | 0.62 | 9358 | 5147 | 6862 |
| 39P/Oterma 1963< | 6.20E+09 | 58 | 0.23 | 3.20 | 0.18 | 3.2E4 | 1.7E4 | 2.3E4 |
| 81P/Wild 2 | 2.09E+10 | 17 | 0.78 | 1.97 | 1.2 | 2200 | 1200 | 1624 |
| 103P/Hartley2 1991[1] | 1.41E+10 | 25 | 0.52 | 0.57 | 7.4<br>δ=0.135 | 85<br>δ=0.02 | 70<br>δ=0.25 | 77<br>δ=0.135 |
| 103P/Hartley2 2010-2011[1] | 2.24E+09 | 160 | 0.084 | 0.57 | 1.2<br>δ=0.135 | 539<br>δ=0.02 | 440<br>δ=0.25 | 485 |
| 2P/Encke 2003 | 8.58E+09 | 42 | 0.32 | 2.55 | 0.30 | 11700 | 6436 | 8580 |
| 2P/Encke 1858 | 1.28E+10 | 28 | 0.47 | 3.20 | 0.28 | 15500 | 8525 | 11367 |
| 9P/Tempel 1 | 1.27E+10 | 28 | 0.47 | 2.75 | 0.38 | 9900 | 5450 | 7270 |
| 45P/H-M-P | 7.95E+09 | 45 | 0.30 | 0.43 | 9.7 | 60 | 33 | 44 |
| 96P/Machholz | 6.55E+09 | 55 | 0.25 | 3.2 | 0.14 | 3.0E4 | 1.7E4 | 2.2E4 |
| 46P/Wirtanen | 4.01E+09 | 90 | 0.15 | 0.60 | 2.5 | 327 | 180 | 239 |
| 28P/Neujmin 1 | 3.58E+09 | 100 | 0.14 | 11.5 | 0.006 | 2.6E06 | 1.4E06 | 1.9E6 |
| 26P/Grigg-Skejelle | 1.64E+09 | 218 | 0.061 | 1.47 | 0.017 | 11700 | 6450 | 8600 |
| 133P/Elst-Pizarro | 1.81E+08 | 1978 | 6.8E-3 | 2.3 | 0.007 | 4.0E05 | 2.2E05 | 2.9E5 |
| 107P/Wilson-Harrin | 2.03E+07 | 1.7E4 | 6.8E-4 | 1.65 | 0.0017 | 1.3E6 | 7E5 | 9.8E5 |
| D/1819W1 Blanpain | 1.40E+07 | 2.6E4 | 5.1E-4 | 0.16 | 0.0012 | 1770 | 970 | 1300 |
| 2006 VW139 | 4.00E+06 | 8.9E4 | 1.5E-4 | 1.8 | 2.8E-4 | 8.8E6 | 4.9E6 | 6.5E6 |
| 3200Phaeton[2](dust) | 4.00E+08 | 895 | 0.015 | 2.5 | 1.4E-2 | ----- | 2.6E5 | 1.7E5 |
| 3200Phaeton[2](dust) | 4.00E+07 | 8950 | 0.0015 | 2.5 | 1.4E-3 | ----- | 2.6E6 | 1.7E6 |
| 3200Phaeton[2](dust) | 4.00E+06 | 89500 | 0.00015 | 2.5 | 1.4E-4 | ----- | 2.6E7 | 1.7E7 |

0. N(comets)= 29.
1. For comet 103P Sanzovo et al. (2010) found 0.02 < δ < 0.25
2. For comet 3200 Phaeton, Li and Jewitt (2013) find only dust ejection.
3. Since comet ISON actually disintegrated RR=1.
4. Since comet C/2002 O4 Hönig actually disintegrated (Sekanina, 2002) RR=1.
5. Except otherwise stated δ = 0.5 has been adopted for most comets.